\documentclass[a4paper,11pt]{article}
\usepackage{graphicx,amssymb,amstext,amsmath}
\usepackage{color}

\definecolor{cbl}{rgb}{0,0,1}                

 \newtheorem{propo}{Proposition}[section]
\newtheorem{rema}{Remark}[section]

\newcommand{\bc}{\begin{center}}
\newcommand{\ec}{\end{center}}
\def\ba#1{\begin{array}{#1}\displaystyle}
\newcommand{\ea}{\end{array}}
\newcommand{\z}{\\[2mm] \displaystyle}

\newcommand{\beq}{\begin{equation}}
\newcommand{\eeq}{\end{equation}}
\newcommand{\beqa}{\begin{eqnarray}}
\newcommand{\eeqa}{\end{eqnarray}}
\newcommand{\no}{\nonumber}
\newcommand{\n}{\nonumber\\}
\newcommand{\bi}{\begin{itemize}}
\newcommand{\ei}{\end{itemize}}

\def\lt#1{\left#1}
\def\rt#1{\right#1}
\def\t#1{\tilde{#1}}
\def\h#1{\hat{#1}}

\def\frc#1#2{\frac{#1}{#2}}

\newcommand{\p}{\partial}

\newcommand{\Pexp}{{\cal P}\exp}

\newcommand{\bra}{\langle}
\newcommand{\ket}{\rangle}
\newcommand{\Z}{{\mathbb{Z}}}

\newcommand{\R}{{\mathbb{R}}}
\newcommand{\C}{{\mathbb{C}}}

\newcommand{\Or}{{\cal O}}

\newcommand{\ep}{\epsilon}

\newcommand{\Tr}{{\rm Tr}}

\newcommand{\End}{{\rm End}}

\newcommand{\Supp}{{\rm Supp}}

\newcommand{\cl}{\overline}

\newcommand{\halmos}{\rule{1ex}{1.4ex}}
\newcommand{\eproof}{\hspace*{\fill}\mbox{$\halmos$}}
\newcommand{\proof}{{\em Proof.\ }}

\newcommand{\tV}{{\t V}}
\newcommand{\ii}{{\tt i}}

\newcommand{\rx}{{\rm x}}
\newcommand{\ry}{{\rm y}}

\begin{document}

\vspace{0.2cm}
\begin{center}

{\Large {\bf Non-equilibrium steady-states in conformal field theory}}

\vspace{0.8cm} 

{\large \text{Denis Bernard${}^{\clubsuit}$}\footnote{{\tt denis.bernard@ens.fr}; Member of CNRS.}
\text{and Benjamin Doyon${}^{\spadesuit}$} \footnote{ \tt{benjamin.doyon@kcl.ac.uk}.}}

\vspace{0.2cm}
${}^{\clubsuit}$ Laboratoire de Physique Th\'eorique\footnote{LPT-ENS is UMR-8549 of CNRS.}, CNRS $\&$ Ecole Normale Sup\'erieure de Paris, France.\\
${}^{\spadesuit}$ Department of Mathematics, King's College London, London, United Kingdom.
\end{center}

\vspace{1cm} 
We present a construction of non-equilibrium steady states in one-dimensional quantum critical systems carrying energy and charge fluxes. This construction is based on a scattering approach within a real-time hamiltonian reservoir formulation. Using conformal field theory techniques, we prove convergence towards steady states at large time. We discuss in which circumstances these states describe the universal non-equilibrium regime at low temperatures. We compute the exact large deviation functions for both energy and charge transfers, which encode for the quantum and statistical fluctuations of these transfers at large time. They are universal, depending only on fundamental constants ($\hbar, k_B$), on the central charge and on the external parameters such as the temperatures or the chemical potentials, and they satisfy fluctuation relations. A key point consists in relating the derivatives of these functions to the linear response functions but at complex shifted external parameters.






\section{Introduction} \label{sectintro}

Triggered by recent experimental advances, the study of quantum systems out of equilibrium is a research line of high current interest, both theoretically and experimentally. Theoretically, this subject is made rather difficult by the diversity of situations it {\em a priori} purports to cover, and by the relative lack of unifying framework extending that of thermodynamics and statistical physics for equilibrium systems. Situations of particular interest, which have the potential of being amenable to a full understanding, are those where steady currents of local quantities exist: steady flows of energy, charge, particles, etc. In these situations, external forces, if any, are constant in time, yet the system is not at equilibrium because there is a permanent creation of entropy. Powerful methods, based on studying fluctuations at large times and their relations with other physical characteristics, have been developed for classical steady states out of equilibrium, and it is hoped that studying fluctuation relations \cite{CohenGal} will lead to an understanding of the fundamental principles of the non-equilibrium physics of steady states (and perhaps of more general non-equilibrium physics).

For quantum systems, the study of steady states, and in particular of fluctuations of the large-time energy/charge/particle transfer, is still very much to be developed. In these situations, the very definition of the higher moments of large-time transfer requires the choice of a measurement protocol: direct or indirect, von Neumann measurements with full wave-function collapses or coupling to an external measuring device with continuous partial collapse. {\em A priori}, results may depend on this choice. Further, as for classical systems, the construction of the steady state itself necessitates an appropriate set-up: external forces or coupling to external reservoirs, the latter being either with Caldeira-Leggett-like models \cite{Cald-Legg}, infinite-time hamiltonian evolution of infinite subsystems (hamiltonian reservoirs) \cite{InfinitReserv}, or non-hamiltonian evolution from integrating out the reservoirs (Lindblad approach) \cite{Lindblad}. Perhaps most importantly, quantum effects themselves may modify our intuition on non-equilibrium physics. The full interplay between these effects and non-equilibrium physics is expected to be seen at quantum critical points: there, quantum fluctuations affect the system macroscopically, so that the dynamical transfer processes should be very much influenced by entanglement and interference.

In the present paper we study non-equilibrium steady states of energy and charge transfer of one-dimensional systems at quantum critical points with dynamical exponent $z=1$, in their low-temperature universal regime. Using the formulation of hamiltonian reservoirs we prove convergence to non-equilibrium steady states; we express these states in a precise fashion through a scattering-matrix formalism in the context of conformal field theory (CFT) and explain their universality; and we obtain exact results for all higher cumulants of large-time transfer of both energy and charge. The set of all higher cumulants is a very important characteristic of a non-equilibrium steady state, sometimes referred to as the full counting statistics (even though the term ``counting'' is more natural in the cases of countable quantities like charges and particles). We gather the cumulants into an exact generating function, which we will refer to as the large deviation function (by a slight abuse of language). From this, we verify that an exact fluctuation relation holds. We also show that the results are largely independent of the measurement protocol, and
we observe that the only characteristics of the universality class needed is the central charge. Finally, we show that there are (at least) two fundamental time scales involved in the approach to the steady state: a short, microscopic (non-universal) time scale, and a larger one related to the speed of propagation of universal excitations. Some of these results for energy transfer were announced in \cite{BD11}, and the present paper provides the details of the calculations, including a description of the algebraic framework for constructing the non-equilibrium steady states, as well as the application of similar ideas to charge transfer. 

Many previous works have studied related quantities in different, less universal situations. The first exact results for the large deviation function of charge transfer in quantum systems, the so-called Lesovik-Levitov formula, was obtained in free fermion systems in the important work \cite{LLformula}. In other works, results have been obtained for critical free boson systems (Luttinger liquids) \cite{GGM}, as well as in some integrable models of quantum field theory \cite{KS,GKLH}. In the original work \cite{LLformula}, the measurement method is indirect, the system being coupled to an external spin which is itself quantum mechanically measured. Other works \cite{xxx,BD-RLM} use direct two-time von Neumann measurement protocols under various detailed prescriptions; the equivalence between all these protocols in the case of charge transfer in free fermion systems is a consequence of these works. Note that sometimes the measurement method is actually implicit, for instance with a physical picture of scattering particles \cite{KS}.

Concerning energy transfer in quantum systems, the first result for large deviation functions was obtained, to our knowledge, in the case of a chain of harmonic oscillators in \cite{OscilChain}. Other studies of non-equilibrium energy currents in the XY model gave the exact steady state and the exact current \cite{AschPill}, based on the $C^*$-algebra mathematical development of \cite{Ruelle}, as well as some aspects of spin correlation functions \cite{AB}. In ref.\cite{Mintchev}, non-equilibrium steady states in Luttinger liquids in star-graph configurations have been constructed and mean energy and charge currents evaluated as well as current-current correlations. In general massive quantum field theory, the exact energy-transfer steady state was described in terms of asymptotic states in \cite{DHouches}.  Further, works in preparation provide various extensions of the present setting: to star-graph configuration, to the quantum Ising model as well as to general massive models \cite{extensions}. 

Our construction is the first application of general CFT principles to the description of flows in steady states out of equilibrium, and the results provide the most universal critical large deviation formula. Since CFT describes large classes of models with interaction (in particular, when the central charge $c$ is not $1/2$ or 1), these are also the first results for energy current and large deviation function beyond models of free fermions or free bosons. With respect to the particular construction of the steady state used, it has been observed that it is a very nontrivial problem to fully describe hamiltonian reservoirs in interacting models (see for instance \cite{BLR}). We believe our results are the first explicit interacting-model results for energy transfers in the context of hamiltonian reservoirs. Finally, we discuss the universality of non-equilibrium steady states, and the time scales involved in its approach.

To deal with conformal field theory out-of-equilibrium using the formulation of hamiltonian reservoirs requires dealing with real time evolution and fields localised at real space positions on the infinite line. This requires adapting some of the usual conformal field theory tools. We devoted part of the paper to propose a framework for such construction. We have tried to make explicit the points which demands a deeper analysis to be fully mathematically rigourous, and we hope that this will motivate the interested readers.

The paper is organized as follows. After a summary of the main results in Section \ref{sectresults}, we present heuristically our set-up and the construction of the non-equilibrium steady state in Section \ref{sectheuristics}. The formal but precise algebraic framework is detailed in Section \ref{sectalgebra}. This allows us to specify in Section \ref{sectness} the non-equilibrium steady state from the construction of an appropriate $S$-matrix. Formula for the large deviation of functions for energy and charge transfer are proved in Sections \ref{sectenergy} and  \ref{sectcharge}. A brief discussion for our results conclude the manuscript.

\section{Overview of results} \label{sectresults}

A standard way of maintaining systems out of equilibrium consists in putting them in contact with two or more different reservoirs, at different temperatures or different chemical potentials for instance. Besides the question as to how to represent the reservoirs, natural questions that one may then try to answer are related to large time behaviour of such systems: Do they reach steady states in the long run? What are these steady states and how do the systems approach them? These are usually difficult questions and tools from $C^*$ algebra or quantum dynamical systems \cite{Davies,Lindblad} have been developed to deal with them. Let us restrict ourselves to the peculiar framework -- general and powerful enough at least for one-dimension systems --  of integrable systems; these we understand as including the universal behaviours of quantum critical points, described by CFT. We may then wonder whether, and how, we could use tools from (equilibrium) integrable models to compute exact  properties of non-equilibrium steady states. Putting an integrable system in contact with standard reservoirs (like Caldeira-Leggett reservoirs \cite{Cald-Legg}) usually breaks integrability, although a few exceptional cases preserving some integrability structures are known \cite{Prosen, Karevski}.  Considering quantum quenches \cite{CalabCardy}, which consists in putting a system out of equilibrium by changing abruptly some external parameters, provides a way to circumvent this difficulty. In these cases, and in particular for local quenches in the Ising chain in transverse magnetic fields \cite{CalabEssler}, it has been proved that expectations of local observables reach steady values, and furthermore, that the system state, viewed as a dual form on the algebra of local observables, converges towards a generalised Gibbs ensemble. This may seem surprising since the dynamics is unitary (there is no external reservoir and the system is closed). The key point is however that convergence only holds at infinite volume and only applies to local observables, so that the system is its own bath (i.e.~the part of the system far away from the local observable locations play the role of effective baths). Note that non-equilibrium quantum quenches are asymptotically generalised Gibbs states, which do not support any fluxes, or energy or charge transfers.

So we choose another framework, preserving integrability, which consists in preparing different copies of some large quantum systems at different thermal equilibrium, say at different temperature and different chemical potential, and then in gluing them at a contact point. As soon as the contact is established, there is going to be energy or charge transfers from one part of the total system to another. Waiting long enough, the steady state should be established -- this is a ``real-time'' construction of the steady state. Similarly to quantum quenches, and since transfers are localised around the contact points, the regions of the different subsystems far away from the contact points play the role of effective reservoirs. Reservoirs represented in this way using hamiltonian evolution of large subsystems at different temperatures or chemical potentials are usually referred to as hamiltonian reservoirs \cite{InfinitReserv, Ruelle, AschPill}; they were first successfully used for heat transfer in \cite{SpohnLebo} in the context of a classical chain of harmonic oscillators. Note that we may view this construction as representing what happens in the system coupled to two large reservoirs a finite distance apart after having zoomed over a bulk region of the quantum system. Under such a zoom, the reservoirs have been scaled away at infinity while preserving the presence of steady currents.

Specifically, we deal with two {\it identical} one-dimensional gapless quantum systems with dynamical exponent $z=1$, whose low energy sectors is described by {\it isomorphic} CFTs. We prepare them at respective temperatures $T_{l,r}$. If these CFTs have $u(1)$ conserved charges, we also prepare them at different chemical potentials $\mu_{l,r}$. We then connect them through a unique contact point, and wait long enough for the steady state to be reached. Energy and charge transfers then take place across the contact point, and these are steady in any finite-size observation domain. Gluing together two isomorphic CFTs may be viewed as modifying the defect at the contact point in a unique bulk CFT, from a purely reflecting defect to the absence of a defect (in fact, our results hold with the presence of any topological defect instead of the absence of a defect).

First we construct the steady state, denoted $\omega_{\rm stat}$, using the algebraic tools of conformal field theory and vertex operator algebra. From large-time evolution with hamiltonian reservoirs, we show that the result can be expressed in the framework of scattering matrix, whose main input is to express the steady density matrix $\rho_{\rm stat}$, dual to the steady state $\omega_{\rm stat}$, in a simple form,
\[ \rho_{\rm stat} = S\, \rho_o\, S^{-1},\]
with $S$ the scattering matrix and  $\rho_o$ the initial density matrix (note that in the following, $\rho_{\rm stat}$ will not be defined with mathematical precision, but $\omega_{\rm stat} = \omega_o\circ {\rm Ad}(S^{-1})$ will).

The next key point is that the steady state factorizes onto right/left movers: right (resp.~left) movers are collectively thermalized at temperature $T_{l}$ (resp.~$T_r$), and chemical potential $\mu_{l}$ (resp.~$\mu_r$) if any. That is,
\[ \omega_{\rm stat}= \sigma_{\beta_l,\mu_l}\otimes \sigma_{\beta_r,\mu_r}.\]
where $\sigma_{\beta,\mu}$ are Gibbs states on the right/left movers with $\beta_{l,r}=1/(k_B T_{l,r})$, and $k_B$ is the Boltzmann constant. This can intuitively be understood as a consequence of the fact that right/left movers have been asymptotically prepared in the far away left/right parts of the subsystem which effectively are at respective temperatures $T_{l,r}$ and chemical potentials $\mu_{l,r}$. It technically follows from the fact that right/left movers have simple scattering in conformal field theories, even in presence of defects, and we explicitly compute the scattering matrix.

We then use this construction to compute the large deviation function for energy and charge transfer. Let $\Delta_tE$ and $\Delta_tQ$ be respectively the energy and charge transferred during a time duration $t$ in the steady regime. Then the large deviation function $F(\lambda,\nu)$ is defined by
\[ F(\lambda,\nu;\beta_{l,r},\mu_{l,r}):=
\lim_{t\to\infty} \frac{1}{t} \log \omega_{\rm stat}\big[\ e^{i\lambda\, \Delta_tE}\; e^{i\nu\, \Delta_tQ}\, \big].\]
The large deviation function codes for all cumulants of the energy and charge transferred\footnote{This is a slight abuse of language: the usual definition of the large-deviation function is as the Legendre transform of this large-time cumulant generating function.}.

This definition is approximate, since, as we mentioned, some care is required to make precise the way the energy or charge transfer is quantum mechanically measured. We will consider both indirect, and direct two-time von Neumann measurement protocols, see Section \ref{sectenergy}. In the case of a direct measurement protocol, one measures the quantity to be transferred at two different times, and takes for $\Delta_tE$ and $\Delta_t Q$ the difference of the values obtained. There is an ambiguity: the first measurement time may be at the ``contact time'', when the hamiltonian reservoirs are connected, or in the established steady state itself, long after the contact time. We prove that the result is the same for the indirect protocol, and the direct two-time measurement protocol with first measurement at the contact time. In fact our results also holds for the charge transfer statistics, but not the energy transfer statistics (contrary to the claim made in \cite{BD11}), in the case of the two-time measurement protocol in the established steady state.

We find a universal formula for the large deviation function, which only depend on universal constants $(\hbar,\, k_B)$ and $T_{l,r}$ and $\mu_{l,r}$. Namely
\[ F(\lambda,\nu;\beta_{l,r},\mu_{l,r})= f(\lambda,\nu;\beta_l,\mu_l)+  f(-\lambda,-\nu;\beta_r,\mu_r),\]
with, for unitary theories,
\[ f(\lambda,\nu;\beta,\mu)= \frac{c\pi}{12\hbar}\big(\frac{1}{\beta-i\lambda}-\frac{1}{\beta}\big) 
 + \frac{\pi(\beta\mu+i\nu)^2}{2\hbar(\beta-i\lambda)} - \frac{ \pi\beta\mu^2}{2\hbar}.\]
The fact that the large deviation function $F$ decomposes as the sum of two terms, each respectively associated to the left/right temperatures and chemical potentials, and thus to the right/left movers, is a direct consequence of the factorisation of the steady state. The function $f$ is a large deviation function in the chiral theory, and as such it is computable using data from the Virasoro or vertex operator algebras.

The above large deviation function satisfies the fluctuation relation:
\[F(i(\beta_r-\beta_l)-\lambda,i(\beta_l\mu_l-\beta_r\mu_r)-\nu;\beta_{l,r},\mu_{l,r}) 
= F(\lambda,\nu;\beta_{l,r},\mu_{l,r}).\]

When chemical potentials are absent, it reduces to that which we gave in \cite{BD11}, in particular the average energy current\footnote{In the case $c=1$ and $T_r=0$, this formula bears similarities with the Stefan-Boltzmann law for the energy radiated by a thermal black body, see \cite{CardyStef}. We thank J. Cardy for pointing out this analogy.} being
\[\bra J_E\ket_{\Delta \beta\neq 0,\Delta\mu=0} = \frac{c\pi}{12\hbar}\, k_B^2(T_l^2-T^2_r).\]
The formula for $F$ shows that energy and charge transfers are correlated in presence of both temperatures and chemical potential differences. This is not surprising if we think of these transfers as emanating from transfers of particles or energy and charge carrier quanta. In particular, a difference of chemical potentials but at identical temperatures induces non zero energy and charge currents, respectively:
\[
\bra J_E\ket_{\Delta\beta = 0,\Delta\mu \neq 0} = \frac{\pi}{2\hbar}\,(\mu_l^2-\mu^2_r),\quad
\bra J_Q\ket_{\Delta\beta = 0,\Delta\mu \neq 0} = \frac{\pi}{\hbar}\,(\mu_l-\mu_r).
\]
Notice that in this case charge transfer fluctuations are gaussian.

Observe also that the case where $\beta_l\mu_l=\beta_r\mu_r$ is particular from the viewpoint of the energy fluctuations, in that it corresponds to the case $\mu_l=\mu_r=0$ up to a shift of the central charge:
\beq \nonumber
	\lt.F(\lambda,0;\beta_{r,l},\mu_{r,l})\rt|_{\beta_l\mu_l=\beta_r\mu_r=:\chi} = \frc{c^*\pi}{12\hbar}\,\lt(\frac{i\lambda}{\beta_l(\beta_l-i\lambda)} - \frac{i\lambda}{\beta_r(\beta_r-i\lambda)}\rt),\quad
	c^* = c + 24\,\chi^2.
\eeq

Besides the factorisation property of the steady state, the main point in the derivation consists in relating the derivative of the (chiral) large deviation function $f$ to the energy $\mathfrak{h}$ and charge $\mathfrak{j}$ current one point functions but at shifted temperature and chemical potential. Namely,
\begin{eqnarray*}
-i\frac{\partial}{\partial\lambda}  f(\lambda,\nu;\beta,\mu)&=&\sigma_{\beta-i\lambda,\frac{\beta\mu+i\nu}{\beta-i\lambda}}\big[ \mathfrak{h}(0)\big] ,\\
	-i\frac{\partial}{\partial\nu}  f(\lambda,\nu;\beta,\mu)&=& \sigma_{\beta-i\lambda,\frac{\beta\mu+i\nu}{\beta-i\lambda}}\big[ \mathfrak{j}(0) \big].
\end{eqnarray*}
This property is one of the consequences of PT-symetry on large deviation functions as we shall explain in \cite{BDnofutur}.

\section{Heuristics: non-equilibrium steady states in CFT}\label{sectheuristics}

The complete description of the non-equilibrium steady state of energy flow in CFT will be provided in Section \ref{sectness}, based on the algebraic setup of CFT developed in Section \ref{sectalgebra}. Here we give the heuristic idea behind it, starting with a general physical description of the setup and obtaining the steady state from standard and simple notions of CFT.

\subsection{Physical description}

We are aiming at describing non-equilibrium states in gapless (critical) one-dimensional quantum systems with dynamical exponent $z=1$, in a setup involving hamiltonian reservoirs. The state is obtained by coupling two such systems which have been independently prepared at different equilibria; in this section we will only consider different temperatures, but the generalization to different chemical potentials is immediate and will be done in Sections \ref{sectcharge} and \ref{sectboth}. For definiteness, we take two identical copies of a gapless quantum system, each of length $R/2$, respectively defined on intervals $[-R/2,0]$ and $[0,R/2]$, and prepared at respective temperatures $T_{l}$ and $T_r$. We then imagine taking $R$ large enough, and connecting the copies through the origin at a large negative time, say $-t_o$, in such a way that, after unitary evolution, the state of the coupled system at time $0$ or later is stationary in any finite observation domain around the contact point. This steady state is out of equilibrium, and, as one can expect and as we will show, there is an energy flow from the high to the low temperature regions. The steady flow takes place on a domain of size of order $v_ft_o$, where $v_f$ is the typical velocity of elementary excitations. This is because, intuitively, it is excitations emitted from the contact point that allow the local state to change from equilibrium to non-equilibrium steady state. A similar physical picture tells us that for the system state to be steady, $v_ft_o$ has to be much smaller than the system size $R$. Otherwise the excitations are going to bounce back on the system boundary walls, thus modifying further the local states; at large times, oscillations would arise. Hence, if $\ell$ is the observation length (the distance to the contact point up to which local observations are being made), the steady state is obtained in the limit $R\gg v_ft_o\gg \ell$. Keeping $\ell$ fixed and finite, this means that the steady state is mathematically defined by the limits $R\to\infty$ and then $t_o\to\infty$ in that order~\footnote{In much of the following we set $v_f=1$, $\hbar=1$, $k_B=1$}. The extreme left and right parts away from the steady-state domain, at a distance much greater than $v_ft_o$ from the contact point, serve as effective thermal reservoirs, each at its own temperature $T_{l,r}$. Thanks to the energy flow between them, it is a simple matter to see that the total thermodynamic entropy of the two asymptotic regions increases.

It is well known that conformal field theory (CFT) describes low-energy equilibrium behaviours of $z=1$ gapless systems (this is not proven but verified in many ways). This is because CFT provides their full asymptotic low-energy sector, arising from the states whose energy differences to the ground state are much lower than microscopic coupling energies $D$. CFT then gives a description of the low-energy thermodynamics, as well as the large-distance correlations of local observables in low-energy states, including, at least it is believed, dynamical correlations. It is important to note that these low-energy behaviours are universal, largely independent of the microscopic structure; CFT can be seen as giving the full quantum mechanics of universal degrees of freedom. It is natural to think that the non-equilibrium steady state of gapless critical systems, obtained from hamiltonian reservoirs as above, will be described similarly, in the region $T_{l,r}\ll D$, by a hamiltonian reservoirs setup in CFT: using a CFT state for the two identical copies initially equilibrated at different temperatures, and using the CFT unitary evolution from this state. Hence our CFT results should indeed give predictions for gapless systems. We will provide conditions for this to be valid in Subsection \ref{ssectuni}.

\subsection{CFT dynamics}

Before being connected, the low-energy sector of the two copies are described by isomorphic CFT models with central charge $c$. After being connected, the total system is again described by a CFT model with identical central charge, but on a segment twice as large. The connection made at the time $-t_o$ can be interpreted in CFT as a change of the point defect at the origin: before contact the defect is totally reflecting, while after contact it is totally transmitting.

Let us recall here a few standard results of CFT. In the bulk, away from boundaries and defects, the energy $h$ and momentum $p$ densities decompose as
\begin{eqnarray}
 h(x,t)=h_+(x,t) +h_-(x,t),\quad
 p(x,t)=h_+(x,t) - h_-(x,t) \label{momdens},
 \end{eqnarray}
with $h_\pm$ the chiral components (right- and left-moving respectively), $(\partial_t\pm\partial_x)h_\pm=0$. This implies local energy conservation, $\partial_t h + \partial_x p =0$.

Assuming that there are no extra degrees of freedom localized at the boundaries or at the contact point, total, but not local, energy conservation holds with boundaries or defects. The boundary and defect conditions we may impose are restricted by this energy conservation. The total energy is the sum of the energy of the left and right parts of the system, that is 
\[ H(t):=\int_{-R/2}^{0^-} dx\, h(x,t) + \int_{0^+}^{R/2} dx\, h(x,t).\]
Demanding energy conservation, that is demanding $\frac{d}{dt} H(t)=0$, and using the bulk local energy conservation law $\partial_t h+\partial_x p=0$ we get:
\[ 0= p(-R/2,t) - [ p(0^-,t)-p(0^+,t)] - p(R/2,t). \]
The topology of our space (a single interval from $-R/2$ to $R/2$) and locality of the densities then forces us to impose $p(-R/2)=0=p(R/2)$ and $p(0^+,t)=p(0^-,t)$. 
The former two conditions are reflecting boundary conditions at the far ends of the two subsystems, namely 
\[ h_+(\pm R/2,t) = h_-(\pm R/2,t),\]
which translate into the usual CFT preserving boundary conditions \cite{Cardy_bdry}. 
The latter condition yields the continuity of the momentum density at the contact point, that is
\begin{eqnarray}\label{bdrycond}
h_+(0^-,t)-h_-(0^-,t)=  h_+(0^+,t)-h_-(0^+,t).
\end{eqnarray}
The dynamics of the uncoupled subsystems and that of the full coupled system correspond to two possible ways of fulfilling this condition.

Let $H_o$ be the hamiltonian before contact. For clarity, we will denote by $h^o_\pm(x,t)$ the time-evolved densities $h_\pm(x)$ under the $H_o$-dynamics. The $H_o$-dynamics corresponds to ensuring (\ref{bdrycond}) by imposing reflecting boundary conditions separately on the left and right subsystems, 
\beq \label{Hobdry}
h_+^o(0^\pm,t)=h_-^o(0^\pm,t).
\eeq
In this case, the system splits into its two independent left and right parts; we have for instance $h_+^o(x,t) = h_+(|x-t|)$ if $R/2>x,t>0$. The hamiltonian $H_o$ is then the sum of commuting left and right subsystem hamiltonians, $H_o=H_o^l+H_o^r$ with  $H^l_o = \int_{-R/2}^{0^-} dx\, h(x)$ and $H^r_o= \int_{0^+}^{R/2} dx\, h(x)$. 

Let $H$ be the hamiltonian of the coupled system. We will denote by $h_\pm(x,t)$ the time-evolved densities $h_\pm(x)$ under the $H$-dynamics. By definition, the $H$-dynamics ensures the relation (\ref{bdrycond}) by imposing continuity for the chiral hamiltonian densities,
\beq \label{Hbdry}
	h_\pm(0^-,t)=h_\pm(0^+,t).
\eeq
In this case, $h_\pm(x,t)$ are both continuous at the contact point, so that, for instance, $h_+(x,t) = h_+(x-t)$ if $R/2>x,t>0$. The hamiltonian is $H = \int_{-R/2}^{R/2} dx\,h(x)$, which formally includes not only the sum of $H^l_o$ and $H^r_o$, as did $H_o$, but also an extra local energy contribution at the point $x=0$ representing the extra link necessary in order to connect the subsystems.

The hamiltonians $H_o=H_o^l+H_o^r$ and $H$ look very similar, differing only by an extra local energy contribution. The correct meaning of this extra contribution is through the dynamics that it generates, rather than as an extra energy term (which has no clear universal meaning: the extra link is an energy term of order $D$, beyond the universal region). That is, the true difference between $H_o$ and $H$ is through the defect condition imposed, eq.(\ref{Hobdry}) or eq.(\ref{Hbdry}), corresponding to an abrupt change of conformally invariant defect. Before the contact time, the defect is factorizing, splitting the system into two parts, while after contact the defect is transmitting, letting the energy flow through. We will clarify this in the algebraic formalism of Section \ref{sectalgebra}. What we are describing can be thought of as a quantum quench \cite{CalabEssler}, but with an initial state being the product of two thermal states (not just the ground state, but a statistical distribution of many low-energy states).

The boundary condition (\ref{Hbdry}) is referred to as a topological defect \cite{JBZ}. It ensures that the chiral hamiltonian densities $h_\pm$ are transmitted through the contact point without reflection. We could as well have considered connecting the two subsystems through an extra defect (like a quantum dot). In this case, we would need the general boundary condition (\ref{bdrycond}), which only guarantees the continuity of the momentum density. It is referred to as non-topological defect. We shall only consider topological defect in the following. Topological defects include homogeneous system with no real defect, but also certain defects making the system non-homogeneous but transparent to the chiral hamiltonian densities. Our results hold for topological defects in general.
 
Note that the phase space is the same for both the $H_o$- and $H$-dynamics: it is made of the fields $h_\pm(x)$, $x\in\R$, and their descendants (discussed in Section \ref{sectness}).

\subsection{S-matrix and steady states}

Recall that we wish to obtain the steady state, at time $0$, by preparing two identical subsystems at different temperatures, coupling them at time $-t_o$ and then taking the large $t_o$ limit. The initial system state is the product of two Gibbs states for the left and right subsystems at respective temperatures $T_{l,r}$. In this heuristic section we will use both the language of states as normalized positive linear form on the algebra of observables (to be specified later) and the dual language of density  matrices. The connection is the usual one, which for the case of the initial state $\omega_o$, or density matrix $\rho_o$, is (where $\beta_{l,r} := T_{l,r}^{-1}$)
\beq \label{rho0}
	\omega_o\big[ \cdots \big] = \Tr\lt(\rho_o \cdots\rt),\quad \rho_o :=\frak{n}\big[e^{-\beta_l H_o^l -\beta_r H_o^r}\big]
	=\frc{e^{-\beta_l H_o^l -\beta_r H_o^r}}{\Tr\lt(e^{-\beta_l H_o^l -\beta_r H_o^r}\rt)}.
\eeq
Here and in the following we use the notation $\frak{n}\big[ \rho \big] := \rho/ {\rm Tr}[\rho]$ to normalize density matrices.

If it exists, the steady-state density matrix $ \rho_{\rm stat}$, dual to the steady state $\omega_{\rm stat}$, is by construction the large-$R$, then large-$t_o$ limit of the initial density matrix evolved with the $H$-dynamics. Since the initial density matrix is $H_o$-invariant we may write this as
\[ \rho_{\rm stat} := \lim_{t_o\to\infty} \lim_{R\to\infty} e^{-it_oH}\, e^{it_o H_o}\, \rho_o\, e^{-it_oH_o}\, e^{it_o H}.\]
This limit exists in an appropriate sense: we will show, essentially, that a similar limit defining $\omega_{\rm stat}$ exists when acting on products of local symmetry fields of the CFT, like the local chiral hamiltonian densities. Note that forgetting about the defect conditions (or the single-link difference between $H_o$ and $H$), the operator $ e^{-it_oH}\, e^{it_o H_o}$ is naively the identity. However, each exponential factor encodes the dynamics on the algebra of observables, and as such, both factors are distinct.  By construction, it is clear that if the limit exists, then $\rho_{\rm stat}$ is invariant under the $H$-dynamics. Let the $S$-matrix be defined formally by
\beq \nonumber
 S:=\lim_{t_o\to\infty} \lim_{R\to\infty} e^{-it_oH}\, e^{it_o H_o}.
 \eeq
It is plain that the $S$-matrix then (formally) intertwines the initial and steady density matrices so that the steady state may (formally) be written as
\beq \label{Srho}
 \rho_{\rm stat}= S\, \rho_o\, S^{-1}.
 \eeq
By duality, steady correlation functions are
\[ \omega_{\rm stat}\big[ \prod_jh_+(x_j)\cdot\prod_k h_-(y_k)\big]:= 
\omega_o\big[\prod_j S^{-1}\, h_+(x_j) \, S\cdot\prod_k S^{-1}\, h_-(y_k) \, S\big],\]
with, for any local operator ${\cal O}$, 
\beq \label{SonO}
S^{-1}\,{\cal O}\,S:=\lim_{t_o\to\infty}\lim_{R\to\infty} e^{-it_oH_o}\, e^{it_o H}\, {\cal O}\, e^{-it_oH}\, e^{it_o H_o}.
\eeq

Let us now compute the $S$-matrix action on fields. By the above definition, it first acts by a forward time evolution with the $H$-dynamics then followed by a backward time evolution with the $H_o$-dynamics. Recall that both dynamics are chiral on operators such as $h_\pm$ in the bulk, and differ only by the defect conditions at the contact point. 

Consider first the action of $S$ on $h_+(x)$. By the $H$-dynamics this is transformed into $h_+(x-t_o)$, for any $R\gg t_o$ large enough so that there is no reflection at the left end point. If $x<0$, so is $(x-t_o)<0$ and the field position never crosses the contact point. When applying the reversed $H_o$-dynamics one never encounters the contact point again and the field goes back unchanged to its original position. Hence $S^{-1}\, h_+(x)\, S= h_+(x)$ for $x<0$. If $x>0$, then $(x-t_o)<0$ for $t_o\gg x$ large enough and the field has crossed the contact point under the $H$-dynamics. Applying next the backward $H_o$-dynamics the field will be reflected at the contact point. Since $h_+$ is reflected in $h_-$ under the $H_o$-dynamics, we get $S^{-1}\, h_+(x)\, S= h_-(-x)$ for $x>0$, so that
\begin{eqnarray}\label{Sonh+}
S^{-1}\, h_+(x)\, S &=& h_+(x),\qquad {\rm for}\ x<0,\\
S^{-1}\, h_+(x)\, S &=& h_-(-x),\quad\; {\rm for}\ x>0. \nonumber
\end{eqnarray}
Similarly, exchanging simultaneously left/right movers with left/right subsystems, we get
\begin{eqnarray}\label{Sonh-}
S^{-1}\, h_-(x)\, S &=& h_-(x),\qquad {\rm for}\ x>0,\\
S^{-1}\, h_-(x)\, S &=& h_+(-x),\quad\; {\rm for}\ x<0.\nonumber
\end{eqnarray}
This defines the action of the $S$-matrix on the hamiltonian densities. Its crucial property is that right movers $h_+$ always end up on the left (real negative axis), while left-movers $h_-$ end up on the right (real positive axis). Using eqs.(\ref{Sonh+},\ref{Sonh-}) and the fact that the initial density matrix $\rho_o$ factorizes on the left and right Hilbert spaces ${\cal H}^l$ and ${\cal H}^r$, it then follows that we have factorization into left- and right-movers in the steady states,
\[ \omega_{\rm stat}\big[ \prod_jh_{+}(x_j)\prod_k h_-(y_k)\big]=\Big(\lim_{R\to\infty} 
\omega_o\big[ \prod_jh_{-s(x_j)}(-|x_j|)\big]\Big)\,\Big(\lim_{R\to\infty} \omega_o\big[ \prod_k h_{-s(y_k)}(|y_k|)\big]\Big),\]
with $s(x):={\rm sign}(x)$. Identifying $h_\pm(-x)$ with $h_\mp(x)$ under $\omega_o$ thanks to the $H_o$-defect conditions, we get
\beq \label{Omfact}
 \omega_{\rm stat}\big[ \prod_jh_+(x_j)\prod_k h_-(y_k)\big]= 
\sigma_{\beta_l}\big[ \prod_j \frak{h}(x_j)\big]\,\sigma_{\beta_r}\big[ \prod_k \frak{h}(y_k)\big],
\eeq
with $\sigma_\beta$ a Gibbs state at temperature $\beta^{-1}$ on the chiral operator $\frak{h}$ representing the hamiltonian densities, cf eq.(\ref{frakh}). This factorization and its derivation is slightly different but similar to that presented in \cite{BD11}.

Note that if $T_l=T_r=T=:\beta^{-1}$, then the density matrix dual to $\omega_{\rm stat}$ is $e^{-\beta H}/\Tr\lt(e^{-\beta H}\rt)$: we recover the equilibrium Maxwell-Boltzmann distribution corresponding to the $H$ dynamics at temperature $T$. We may write $H=H_++H_-$, where $H_+$ performs the time evolution of right-movers (and commutes with left-movers), and $H_-$ that of left-movers (and commutes with right-movers). Then it is clear that the stationary state can be represented by
\beq\label{stathphm}
	\omega_{\rm stat}\big[\cdots\big] = \Tr\lt(\frak{n}\big[e^{-\beta_l H_+ - \beta_r H_-}\big]\cdots\rt)
\eeq
This is in agreement with the general understanding that in quantum field theory, the steady-state density matrix is an exponential of this form where $H_+$ measures the total energy of asymptotic right-movers, and $H_-$ that of asymptotic left-movers \cite{BD11,DHouches}. The fact that this state is steady, i.e.~$H$-invariant, is immediate to check: the $H$-dynamics is left/right-moving on $h_\pm(x)$ for all $x\in\R$, and $\sigma_\beta$ is translation invariant. Tracing back the steps, the $H$-invariance of $\omega_{\rm stat}$ is a consequence of the $S$-matrix action (\ref{Sonh+},\ref{Sonh-}) combined with the identification $h_\pm(-\epsilon)\equiv h_\mp(\epsilon)$ for $\epsilon\to 0$, that holds under $\omega_o$. Finally, note that (\ref{Omfact}) holds for any chiral descendants of the hamiltonian densities $h_\pm(x)$.

The previous argument applies to chiral hamiltonian densities and their descendants for any topological defect. It applies because, by definition, a topological defect is such that it intertwines these chiral fields on both side of the contact point. This intertwining property only applies to the chiral hamiltonian densities and its descendants. Commutation relations between the topological defect and other chiral operators may be more involved -- and they are model-dependent details. However, if the defect is trivial, i.e. the identity so that the coupled system is homogeneous, the previous argument applies and the steady state measure factorizes on any chiral operators. In fact, it is clear that this can be generalized to any chiral current in CFT, for instance the $U(1)$ current associated to charge transfer, provided the defect is topological for the corresponding chiral algebra.

In the next section, we will put these heuristic arguments on a more algebraically precise basis. The factorization properties eq.(\ref{Omfact}) will be instrumental.

Finally, we note that formula (\ref{Omfact}) and its derivation only strictly hold for local operators located at nonzero distances away from the contact point (i.e.~all $x_j\not=0$ and all $y_k\not=0$). This is first because the point $x=0$, of course, needs a refinement of the initial state $\omega_o$. But second, and more importantly, the renormalization of fields at the point 0 is incompatible with the $S$-matrix action; for instance, the descendant field $h_+(0^+)h_+(0^-)$ is well defined under $\omega_o$ but not under $\omega_{\rm stat}$, as an extra renormalization is necessary in order to take care of the zero distance between the two fields involved. The fundamental reason behind this is that the $S$-matrix does not preserve the full operator algebra. We will provide a precise statement of this subtle point in Section \ref{sectness}, using the algebraic setup of the next section. This observation also relates to the applicability of the CFT steady state as a universal low-energy limit of quantum-chain steady states, a point which we discuss in Subsection \ref{ssectuni}.

\section{CFT as an algebra of local fields and topological defects}\label{sectalgebra}

A precise formulation of the state $\omega_{\rm stat}$ and of the $S$-matrix discussed above, as elements acting on an algebra of CFT fields, necessitates a precise formulation of certain CFT notions. Here we introduce such a formulation based on the description of CFT in terms of vertex operator algebras, combined with algebraic-QFT-like ideas of local fields. The main point is to keep real time evolutions along with fields at real space positions, instead of using formal variables of vertex operator algebra and imaginary time evolution. We will provide a description where the limit $R\to\infty$ has already been taken.

From general QFT notions, we know that local fields cannot be multiplied if they are at coincident positions\footnote{A non-associative ``normal-ordered'' product may be defined in CFT, but we will not need this here.}. Hence we use the notion (found, e.g., in the theory of unbounded operators) of partial algebras: the multiplication exists if and only if the positions of local fields are non-coincident, and we generate our basic space from this condition. On this space, we then define states as appropriate linear functionals. This description does not contain the information about the {\em topology} of the space on which the system lies, so we must provide additional structures to encode this information. We may expect, for instance, the additional structure of the state $\omega_o$ to be encoding the topology of two semi-lines $\R_+$ and $\R_-$ (recall that $R=\infty$), while that of the state $\omega_{\rm stat}$ to encode the topology of the line $\R$. This additional structure, which we will refer to as a {\em local structure}, is provided by the operator product expansions (OPEs). These are equalities, in the weak sense with respect to the state, relating products of two local fields at different positions but near enough to each other, with infinite (graded) series of local fields at a single position. The OPEs can be seen as giving a nontrivial algebraic structure to our algebra (whose ``associativity'' is a consequence of the theory of vertex operator algebras). Given the local structure, one can then discuss smooth time-evolutions: one-parameter groups of automorphisms of the algebraic structure.

We will consider a canonical algebra whose local structure encodes the topology associated with the purely transmitting defect, with canonical state $\sigma_{\beta_l,\beta_r}$ and time evolution $U_t$ where $h_\pm$ are right/left-movers, respectively, on the whole line. From this, an isomorphism $\Phi_o$ is constructed, which implements the purely reflecting defect by composition, see (\ref{uto}) and (\ref{omegao}). Naturally, since the $S$-matrix intertwines the $H_o$ and $H$ dynamics, it will turn out to be exactly equal to $\Phi_o$. We will see that the $S$-matrix is an algebra automorphism for a weaker local structure than that of either $\omega_o$ or $\omega_{\rm stat}$.

\subsection{Vertex operator algebras and Virasoro algebra}

In the standard description of conformal field theory via vertex operator algebras (see for instance \cite{vertex-op}), a model of CFT is characterized by a choice of a vertex operator algebra $V$ (the symmetry algebra of the theory), and a choice of a module ${\cal M}$ for $V$ (or for the tensor product $V\otimes V$). Every vertex operator algebra contains the Virasoro algebra, which has basis $L_n:n\in\Z$ and commutation relations
\[
	[L_m,L_n] = (m-n)L_{m+n} + \frc{c(m^3-m)}{12} \delta_{m,n},
\]
and ${\cal M}$ is naturally also a module for the Virasoro algebra. Further, a vertex operator algebra is naturally a module for itself, and ${\cal M}$ always contains a submodule isomorphic to $V$ (this is the identity module for $V$). The vertex operator algebra $V$ represents the algebra of chiral symmetry currents, and includes, in particular, the ``identity field'' ${\bf 1}\in V$. 

The vertex operator map $Y(\cdot,\rx)$ is a map from $V$ to formal series ${\rm End}(V)[[\rx,\rx^{-1}]]$ (series in the formal variable $\rx$ with coefficients that are endomorphisms of $V$): $Y(v,\rx) = \sum_j \rx^{-j-1} v_j$ where $v_j\in{\rm End}(V)$ are referred to as the modes of the element $v$. The sum is implicitly over all integers $j\in\Z$. There is a particular element (the normalization is for later convenience)
\beq\label{frakh}
	\frak{h} = -\frc1{2\pi} L_{-2}{\bf 1}
\eeq
whose normalized modes satisfy the Virasoro algebra, $(-2\pi) Y(\frak{h},\rx) = \sum_j \rx^{-j-2} L_{j}$. The formal series $Y(v,\rx)w$ reproduces the operator product expansion of the two CFT fields $v$ and $w$ at a (formal) distance $\rx$. See \cite{vertex-op} for a full definition of vertex operator algebras, including commutativity and associativity.

The algebra $V$ is graded by $L_0$, and in the context of vertex operator algebras, one has $V = \coprod_{j \in \mathbb{Z}} V^{(j)}$ (direct sum, or coproduct) where $j$ are $L_0$-eigenvalues and where every $V^{(j)}$ is finite dimensional.

In the present work, we depart from some of the usual vertex operator algebra notions in two ways. First, we allow graded infinite series $\sum_{j\in\Z} v^{(j)}$ for $v^{(j)}\in V^{(j)}$ (that is, we are effectively looking at the direct product $\prod_{j \in \mathbb{Z}} V^{(j)}$, instead of the direct sum). We will denote the resulting vector space by $\cl{V}$, where graded series have been adjoined linearly. We note that for $v\in V$ and $j\in\Z$, the endomorphism $v_j$ naturally acts on $\cl V$. Dealing with graded series requires that either a finite number of operations occur grade by grade, or that, if an infinite number of operations occur, the result be convergent in an appropriate topology. We will come back to such issues when they appear below. Second, we consider vertex operators $Y(v,x)$ with real variables $x\in\R^\times:=\R\setminus \{0\}$ instead of formal variables $\rx$. For instance, the series $Y(v,x)w$, for $v,w\in V$, $x\in\R^\times$, is a graded series instead of being a formal series, hence it lies in $\cl{V}$ but not in $V$.

\subsection{Algebras of local fields $\tV$ and $\h V$} \label{algebra} \label{topo}

Here and below, we fix $V$ to be a vertex operator algebra as above. Consider elements of the form $v(x)$ for $v\in V$ and $x\in \R$ and let $V(\R)$ be the graded vector space spanned by these. For every $x\in\R$, the map $V \to V(\R): v\mapsto v(x)$ is linear and grading preserving, and ${\bf 1}(x)={\bf 1}$. An element $v(x)$ will be interpreted as the field at positions $x\in\R$ associated to the vector $v$. We form the linear algebra $V^\sharp$ spanned by all free products (juxtapositions) of elements in $V(\R)$ under the relations making ${\bf 1}$ a unit. The space $\tV$ is the subspace spanned by products with non-coincident positions under the additional relation of commutativity $\xi\zeta\sim\zeta\xi$, that is, $\tV={\rm span}\{v_1(x_1)\cdots v_n(x_n):v_i\in V,\,x_i\in\R\;\forall\;i,\;x_i\neq x_j\;\forall\;i\neq j,\;n\geq 1\}/\sim$. This makes $\tV$ into a partial algebra, where the algebra product of $\xi$ with $\zeta$, which exists only if the fields in $\xi$ are at different positions from those of all fields in $\zeta$, is the juxtaposition, and where the order of the factors is unimportant. Connecting with the previous heuristic considerations, the physically relevant algebra (without considering the local structure yet) will be $\tV\otimes \tV$, with the interpretation that the first factor describes right-movers and the second, left-movers, as is usual in CFT. In particular, the energy densities $h_\pm(x)$ are associated with the vertex operator algebra element $\frak{h}$,  eq.(\ref{frakh}),
\beq\label{hphm}
	h_+(x) := \frak{h}(x)\otimes {\bf 1},\quad
	h_-(x) := {\bf 1}\otimes \frak{h}(x).
\eeq

Given a vector space $W$, we may make it a {\em supported vector space} by assigning to every $\zeta\in W$ a set $\Supp(\zeta)$ of closed subsets of $\R$, the supports of $\zeta$, such that if $A\in\Supp(\zeta)$ then $B\in\Supp(\zeta)$ for every $B\supset A$, and such that for every $\zeta,\xi\in W$ and $a\in\C$, if $A\in\Supp(\zeta)$ and $B\in\Supp(\xi)$ then $A\cup B\in \Supp(\zeta + \xi)$, and $A\in \Supp(a\zeta)$. A partial algebra ${\cal A}$ is consistent with its supported vector space structure if for every $\zeta,\xi\in  {\cal A}$ such that there are $A\in \Supp(\zeta)$ and $B\in\Supp(\xi)$ satisfying $A\cap B = \emptyset$, there exists a product $\zeta \xi\in {\cal A}$, and if for every $A\in\Supp(\zeta)$ and $B\in\Supp(\xi)$, $A\cup B\in\Supp(\zeta\xi)$ if the product exists. We see that the partial algebra $\tV$ is consistent with the support structure generated by $\{x\}\in\Supp(v(x)):v\in V,x\in\R$; for instance, we then see that the set $\{x_1,\ldots,x_n\}$ is a support of $v_1(x_1)\cdots v_n(x_n)$.

We now extend the algebra $\tV \stackrel{\rm ext}\mapsto \tV^{\rm ext}$. The basic point is that we also want to have graded series, elements of $\cl{V}$, at positions $x$, in order to account for OPEs.  Hence let $\cl V(\R)$ be the vector space $V(\R)$ to which we linearly adjoin all graded series of the form $\sum_{j\in\Z} v^{(j)}(x)$ for $v^{(j)}\in V^{(j)}$ and $x\in\R$. We identify this canonically and linearly with the space spanned by elements of the form $v(x)$ for $v\in\cl V$ and $x\in\R$, so that $\sum_{j\in\Z} v^{(j)}(x) = \big(\sum_{j\in\Z} v^{(j)}\big)(x)$.

But this space is too big for our purposes. We consider a subspace of it, and we give it a support structure, denoting the supported space by $\cl{V}_{\rm OPE}(\R)$. We generate $\cl{V}_{\rm OPE}(\R)$ starting from the smaller $V(\R)$ by essentially adjoining in a recursive fashion the graded series, in $\cl V(\R)$, corresponding to the operator product expansions. We do it as follows. First, every graded series $\big(Y(v,x)w\big)(y)$, for $v,w\in V$, $x\in\R^\times$ and $y\in\R$, is in $\cl{V}_{\rm OPE}(\R)$, and we assign to it the supports generated by the interval $[y-|x|,y+|x|]$. We then adjoin other graded series in $\cl{V}_{\rm OPE}(\R)$ by similarly using the OPE with all elements already there, and repeat the process, so that we have nested OPEs. There is an additional subtlety, in that in order to do so, we need to sum infinite sequences of graded series, such that there may be infinitely many terms at each grade. For this, we use, for the coefficient of every basis vector, the topology of convergence for complex numbers (of course this procedure is independent of the choice of basis). So, we say that if $\zeta=\sum_{j\in\Z} v^{(j)}(x)$ and $\xi=\sum_{k\in\Z} w^{(k)}(y)$ are both in $\cl{V}_{\rm OPE}(\R)$ and if they have disjoint supports, then the infinite sum of graded series $\sum_{j,k\in\Z} \big(Y(v^{(j)},x-y)w^{(k)}\big)(y)$ is in $\cl{V}_{\rm OPE}(\R)$ and it has supports generated by the intervals $[y-d_A,y+d_A]$, $d_A = {\rm max}_{x'\in A}(|x'-y|)$ for all $A\in \Supp(\zeta)$. Of course, a crucial point here is that the infinite sequences of graded series should sum to graded series with finite coefficients. This is guaranteed, recursively, by the condition on $\zeta$ and $\xi$, above, having {\em disjoint supports}, and then by the specific choice of supports for the resulting graded series.

The supported partial algebra $\tV^{\rm ext}$ is then the free commutative algebra over $\cl{V}_{\rm OPE}(\R)$ that is consistent with its support structure.

Finally, we put the nontrivial structure of operator product expansions, giving the algebra its local structure: the resulting algebra $\h V$ is obtained by imposing on $\tV^{\rm ext}$ the condition that the operator product expansion holds. In its general form, this means that if $\zeta=\sum_{j\in\Z} v^{(j)}(x)$ and $\xi=\sum_{k\in\Z} w^{(k)}(y)$ are both in $\cl{V}_{\rm OPE}(\R)$ and if they have disjoint supports, then
\beq\label{OPE}
	 \zeta\xi \stackrel{OPE}\sim
	 \sum_{j,k\in\Z} \big(Y(v^{(j)},x-y)w^{(k)}\big)(y).
\eeq
In particular,
\[
	v(x)w(y) \stackrel{OPE}\sim \big(Y(v,x-y)w\big)(y),
\]
for all $v,w\in V$ and $x,y\in\R$ with $x\neq y$. That is, $\h V = \tV^{\rm ext}/\stackrel{OPE}\sim$. This makes $\h V$ into a commutative and associative algebra. By our construction, the condition \eqref{OPE} guarantees that we can describe the full algebra $\h V$ as a span of elements of the form
\beq\label{basishV}
	v_1(x_1)\cdots v_n(x_n):v_i\in V,\,x_i\in\R\;\forall\;i,\;x_i\neq x_j\;\forall\;i\neq j,\;n\geq 1.
\eeq
Indeed, this is because any graded series in $\h V$ comes from using, recursively, the OPEs. In fact, like for the algebra $\t V$, these elements also form a basis of $\h V$.

\begin{rema}\label{rema1}
Of course, a crucial point in the recursive construction of $\cl{V}_{\rm OPE}(\R)$ is that the infinite sequences of graded series should sum to graded series with finite coefficients, which is a consequence of the choice of supports. Although we have not developed the full details, we expect a complete proof of this statement to be possible within the framework of vertex operator algebras -- note that we are dealing with the simplest part of the theory of vertex operator algebras: the algebra itself, not its modules and intertwiners. The idea behind our expectation is that in a CFT correlation function, the OPE between two fields $v(x)w(y)$, as a series expansion in $x-y$ with coefficients that are fields at $y$, is expected to be a convergent series whenever $|x-y|$ is smaller than the distance between $y$ and the position of any other field in the correlation function. The resulting series can then be seen as lying on the interval centered at $y$ of length $2|x-y|$. Then, we can apply the OPE term by term for two such series, $\zeta$ and $\xi$, in a similar way, whenever their regions of convergence (the intervals where they lie) are disjoint. The result is a graded series that can be seen as lying on an interval centered at the center of the region of convergence of $\xi$ and long enough so that it covers the region of convergence of $\zeta$ (and then it automatically covers that of $\xi$). This is exactly what we implement with the above prescription on the supports.
\end{rema}

\begin{rema}\label{rema2}
There are two other subtle points for which we have not provided proofs, but which we expect follow form the theory of vertex operator algebras: (1) the fact that the equivalence relation (\ref{OPE}) leads to a commutative and associative algebra product (this should be an immediate consequence of the commutativity and associativity properties of vertex operator algebra); (2) the fact that the elements \eqref{basishV} form a basis for $\h V$.
\end{rema}

Our construction $\tV \to \tV^{\rm ext} \to \h V$ then naturally gives a map
\beqa \label{qR}
	\ii^\vee\;:\; \tV &\to& \h V\n
	v_1(x_1)\cdots v_n(x_n) &\mapsto&
	v_1(x_1)\cdots v_n(x_n)
\eeqa
(where on the right-hand side we implicitly take the coset). Thanks to our comments above about commutativity and associativity of the algebra $\h V$ and about its basis, we see that this map is trivially an algebra isomorphism.

The construction (\ref{qR}) induces maps, both denoted $q^\vee$ for simplicity, on linear functionals $\in \tV^*$ and on endomorphisms\footnote{An endomorphism $\t K\in\End(\tV)$ contains the information of a map $\t K_{\Supp}$ between subsets of $\R$ such that if $\t K(\zeta) = \xi$ then $\t K_{\Supp}(A)  \in \Supp(\xi)$ for all $A\in\Supp(\zeta)$.} $\in {\rm End}(\tV)$ of $\tV$. The image of a linear functional or an endomorphism under $q^\vee$ is an object defined by extending linearity to graded series, and by acting on cosets in the natural way (that is, element by element). In general, this process does not give rise to a linear functional on $\h V$ or to an endomorphism of $\h V$: the set of cosets may not be preserved. We will say that $\t\omega\in \tV^*$ is compatible with $\h V$ if $q^\vee(\t\omega)\in \h V^*$, and that $\t K\in {\rm End}(\tV)$ is compatible with $\h V$ if $q^\vee(\t K)\in {\rm End}(\h V)$. In these cases, we have
\beq\label{qveeomegaK}
	q^\vee(\t\omega) = \t\omega\circ (\ii^\vee)^{-1},\quad 
	q^\vee(\t K) = \ii^\vee\circ \t K \circ (\ii^\vee)^{-1}
\eeq
(recall that $\ii^\vee$ is an isomorphism of algebras, so its inverse acts on $\h V$). If $\t K_1$ and $\t K_2$ are both compatible with $\h V$, then their product is and we see that
\beq\label{comp}
	q^\vee(\t K_1 \t K_2) = q^\vee(\t K_1)q^\vee(\t K_2)
\eeq
so that, in particular, $q^\vee$ preserves the algebra of $\tV$-automorphisms compatible with $\h V$.

If $\t K$ is a $\tV$-automorphism but is {\em not} compatible with $\h V$, then we may still make sense of the map $K:=q^\vee(\t K)$ by acting on each element in each quotient. This maps quotients to a different collection of subsets, which gives a new algebra $\h V_K$:
\beq\label{qRK}
	\h V_K:=K(\h V).
\eeq
Since $\t K$ is a $\tV$-automorphism, then as before, this construction induces an isomorphism
\beqa
	\ii^\vee_K\;:\; \tV &\to& \h V_K\n
	v_1(x_1)\cdots v_n(x_n) &\mapsto&
	v_1(x_1)\cdots v_n(x_n)\label{iK}
\eeqa
where on the right-hand side we implicitly take the associated subset in $\h V_K$ (it is unique and it exists). The map $K$ has the explicit expression
\beq\label{qK}
	K = \ii^\vee_K \circ \t K \circ (\ii^\vee)^{-1}.
\eeq
Note that if $\t K$ is compatible with $\h V$, then in particular $\ii^\vee_K = \ii^\vee$, and we recover the previous expression.

When $\t K$ is not compatible with $\h V$, then $\h V_K$ has a {\em different} local structure than that of $\h V$. This local structure is implemented by the isomorphism $\ii_K^\vee$ defined in (\ref{iK}), and by the maps $q_K^\vee$ on linear functionals and endomorphisms, defined again by extending and taking the quotient in the natural way according to $\h V_K$. Note that if the endomorphism $\t L$ is compatible with $\h V$, then the endomorphism $\t L_K:=\t K \circ \t L\circ \t K^{-1}$ is compatible with $\h V_K$; similarly if $\t\omega$ is compatible with $\h V$, then $\t\omega_K:=\t\omega\circ \t K^{-1}$ is compatible with $\h V_K$. Then, using equations similar to (\ref{qveeomegaK}) but for the case of $\h V_K$ (in particular with $\ii_K^\vee$ instead of $\ii^\vee$), and using (\ref{qK}), we may express the maps $q_K^\vee$ in terms of the maps $q$:
\beq
	q_K^\vee(\t \omega_K)=q(\t\omega)\circ K^{-1} ,\quad
	q_K^\vee(\t L_K)=K\circ  q^\vee(\t L)\circ K^{-1}.
	\label{qKmap}
\eeq
We will use these concepts below in order to discuss defects.

\subsection{A Gibbs state and time evolution on $\h V$} \label{Sect:Gibbs}

We now provide on $\h V$ a family of linear functionals $\sigma_\beta$, $\beta>0$, which are Gibbs states defined via the nonzero-temperature correlation functions of CFT.

We first choose the module ${\cal M}$ simply to be the identity module of the vertex operator algebra $V$. This module is endowed with an inner-product $(\cdot,\cdot)$ that gives the usual correlation functions of CFT in the planar geometry (i.e.~infinite-length chains in their ground states): 
\beq\label{zeroT}
	\iota_{\{\rx_i\mapsto x_i\}} \big({\bf 1}, Y(v_1,\rx_1)\cdots Y(v_n,\rx_n)){\bf 1}\big)
\eeq
for $v_j \in V, \,x_j\in \R\subset \R^2$. Here $\iota_{\{\rx_i\mapsto x_i\}}$ maps the formal series $\big({\bf 1}, Y(v_1,\rx_1)\cdots Y(v_n,\rx_n)){\bf 1}\big)$ to an analytic function of $x_1,\ldots,x_n$ (whose only singularities are poles) by analytically continuing from the region of convergence $x_1>\cdots >x_n$.

In CFT, correlation functions at nonzero temperature $T$ can be described via a map $y\mapsto x$ from $\C^\times = \C\setminus \{0\}$ to the cylinder ${\rm Im}(x)\in[0,\beta)$, with $\beta = 1/T$; the map is given by $x= (2\pi T)^{-1}\log y$. This map induces a map $v\mapsto v^{(\beta)}$ the image of which forms an isomorphic vertex operator algebra under a modified $Y$-map $Y[\cdot,y] = Y(y^{L_0}\cdot,y-1)$ \cite{vertex-op-cyl}. Hence, for $v_k\in V,x_k\in \R:\;k=1,\ldots,n$, we define $\t\sigma_\beta\in\t V^*$ by
\beq\label{cfT1}
	\t\sigma_\beta\big[v_1(x_1) \cdots v_n(x_n)\big]
	= \lt[\iota_{\{\ry_i\mapsto y_i\}} \big({\bf 1}, Y({\ry_1}^{L_0}v_1^{(\beta)},\ry_1)\cdots Y(\ry_n^{L_1}v_n^{(\beta)},\ry_n){\bf 1}\big)\rt]_{\big\{y_k = e^{\frc{2\pi x_k}\beta}\big\}}.
\eeq
We note in particular that (see (\ref{frakh}) and note the normalization of $\frak{h}$)
\beq\label{transfoh}
	\frak{h}^{(\beta)} = \big(\frc{2\pi}{\beta}\big)^2\big(\frak{h}+\frc{c}{48\pi}{\bf 1}\big).
\eeq
By our construction and in particular Remarks \ref{rema1} and \ref{rema2}, $\t\sigma_\beta$ is compatible with $\h V$ and
\beq\label{defsigma}
	\sigma_\beta := q^\vee \lt(\t\sigma_\beta\rt) \in \h V^*.
\eeq

There is in fact another way of describing the same state, which will be useful below. Correlation functions at nonzero temperature are also traces on appropriate modules. By standard arguments, CFT for a quantum system of a finite (scaled) length $R/2$ can be described using a module ${\cal M}$ for a single copy of $V$ describing states on a circle of circumference $R$ (this is because thanks to reflections at the boundaries, describing only right-movers is sufficient). The state of nonzero temperature $T$ is geometrically a torus with cycles of lengths $R$ and $\beta=1/T$, obtained by tracing over the module ${\cal M}$:
\beq\label{precfT2}
	\lt[\iota_{\{\ry_i\mapsto y_i\}} \Tr_{\cal M}\lt( \frak{n}\lt[e^{-\frc{2\pi \beta}{ R}L_0}\rt]Y(\ry_1^{L_0}v_1^{(R)},\ry_1)\cdots Y(\ry_n^{L_0}v_n^{(R)},\ry_n)\rt)
	\rt]_{\big\{y_k = e^{\frc{2\pi i x_k} R}\big\}}
\eeq
Here the element  $L_0$ of the Virasoro algebra is involved. The infinite-$R$ limit then gives the state of nonzero temperature on the line:
\beqa\label{cfT2}
	\lefteqn{\t \sigma_\beta\big[v_1(x_1) \cdots v_n(x_n)\big] } &&\n &=&
	\lim_{R\to\infty}
	\lt[\iota_{\{\ry_i\mapsto y_i\}} \Tr_{\cal M}\lt( \frak{n}\lt[e^{-\frc{2\pi \beta}{R}L_0}\rt]Y(\ry_1^{L_0}v_1^{(R)},\ry_1)\cdots Y(\ry_n^{L_0}v_n^{(R)},\ry_n)\rt)
	\rt]_{\big\{y_k = e^{\frc{2\pi i x_k} R}\big\}}.
\eeqa
The fact that this gives the same state as \eqref{cfT1} is a consequence of modular invariance of ${\cal M}$ and unitarity\footnote{For non-unitary models, \eqref{cfT2} is the appropriate finite-temperature correlation function.}.

We mention that from the data of the one-point functions $\sigma_\beta\big[v(x)\big]$ (which are in fact independent of $x$), it is possible to deduce the correlation functions $\sigma_\beta\big[v_1(x_1) \cdots v_n(x_n)\big]$ by seeing them as giving values on the real subspace $\R^n$ of functions on $\C^n$ with appropriate analytic properties. These properties include periodicity along the imaginary directions with period $\beta$, and the conformal Ward identities, specifying poles at colliding positions $x_i=x_j$ of a fixed form determined by the vertex operator algebra.

There is a natural time evolution associated to the Gibbs states defined above, which preserves the state. Let $\t U_t^\vee:t\in\R$ be a one-parameter group of $\tV$-automorphisms defined by
\beq\label{Utvee}
	\t U_t^\vee(v(x)) = v(x-t),\quad v\in V,\,x\in\R.
\eeq
These automorphisms are compatible with $\h V$ and we define
\beq\label{defut}
	U_t^\vee := q^\vee\lt(\t U_t^\vee\rt) \in {\rm End}(\h V)
\eeq
which, thanks to (\ref{comp}), form a one-parameter group of automorphisms of $\h V$. Physically, the compatibility indicates that $ U_t^\vee$ represents a time evolution in agreement with the topology on the line, here for a single copy of $\h V$, and the definition (\ref{Utvee}) means that this single copy is right-moving. These automorphisms preserve the Gibbs state $\sigma_\beta$ defined above: $\sigma_\beta \circ U_t^\vee = \sigma_\beta$.

\subsection{Canonical topology, states and time evolution on the physical algebra ${\cal V}=\h V\otimes \h V$}

The physically relevant algebra is $\tV\otimes \tV$. This is naturally a supported vector space (with a slightly extended notion of support), where supports are generated in the natural way from those of $\tV$  and lie on the disjoint union $\R\cup_{\rm disj} \R$. We now consider on $\tV\otimes \tV$ a canonical local structure, canonical states and a canonical time evolution.

The canonical local structure corresponds to performing the construction (\ref{qR}) on both factors, leading to maps $q= q^\vee \otimes q^\vee$ and $\ii = \ii^\vee\otimes \ii^\vee$. We will denote the corresponding algebra by
\[
	{\cal V}:=\h V\otimes \h V.
\]
Canonical states are defined by factorization:
\beq\label{sigma2}
	\sigma_{\beta_1,\beta_2} = \sigma_{\beta_1} \otimes \sigma_{\beta_2} = q(\t\sigma_{\beta_1,\beta_2}),\quad\beta_1,\beta_2>0 
\eeq
where $\t\sigma_{\beta_1,\beta_2} := \t\sigma_{\beta_1} \otimes \t\sigma_{\beta_2}$ (these are compatible with $\cal V$). Finally, the canonical time evolution is the one-parameter group $U_t:t\in\R$ of $\h V\otimes \h V$-automorphisms defined by
\beq\label{ut}
	U_t= U_t^\vee\otimes U_{-t}^\vee = q(\t U_t)
\eeq
where $\t U_t := \t U_t^\vee\otimes \t U_{-t}^\vee$ (compatible with $\cal V$). In terms of our heuristic discussion above, where the hamiltonian $H$ is involved, the time evolution $U_t(\cdots)$ is interpreted as the conjugation $e^{iHt}\cdots e^{-iHt}$; it physically represents the evolution on the line, where in ${\cal V}=\h V\otimes \h V$ (hence also in its restriction $\tV\otimes \tV$), the first factor is right-moving and the second is left-moving. In particular, this time evolution, along with the correspondence (\ref{hphm}), implies the usual time evolution for right- and left-moving energy densities,
\beq\label{hpmxt}
	h_\pm(x,t) := U_t(h_\pm(x)) = h_\pm(x\mp t).
\eeq
The automorphisms (\ref{ut}) preserve the linear functionals $\sigma_{\beta_1,\beta_2}$,
\beq\label{sym}
	\sigma_{\beta_1,\beta_2} \circ U_t = \sigma_{\beta_1,\beta_2}.
\eeq

The topology represented by the local structure of $\cal V$ is that where both right- and left-movers on separate lines, so that the full quantum system, containing both types of movers, is on the line. Note however that $\sigma_{\beta_1,\beta_2}$ is not in general a physical Gibbs state for the corresponding quantum system, except when $\beta_1=\beta_2=\beta$, where we then have a state of nonzero temperature $\beta^{-1}$ on the line (which we do not make use of in this paper).

\subsection{Boundaries and topological defects} \label{sect:bdry}

In our heuristic discussion we introduced two types of defects: a reflecting defect at $x=0$ and a topological defect, effectively the absence of any defect from the viewpoint of symmetry currents. These two types of defects were associated with particular time evolutions. Also, on systems with the former type of defect we put a product of Gibbs states where both halves are thermalized independently at temperatures $\beta_l$ and $\beta_r$. On systems with the latter type of defect, we had instead a non-equilibrium steady state. We now describe how to implement these defect conditions on the local field algebra $\tV\otimes \tV$, both from the viewpoint of the correct local structure and of the time evolution, and we explain how to implement the initial product of Gibbs state on the former type of defect. We will describe the non-equilibrium steady state in the next section.

The main idea is to intertwine the physical degrees of freedom in the algebra $\tV\otimes \tV$ in such a way that the local structure $\ii$, the canonical time evolution $U_t$ and the canonical state $\sigma_{\beta_1,\beta_2}$ give rise to the physical local structure, time evolution and state. The intertwining of the physical degrees of freedom will be implemented by a $\tV\otimes \tV$-automorphism $\t\Phi$,
\[
	\t\Phi:\tV\otimes \tV \to \tV\otimes \tV.
\]
As in (\ref{qRK}), there is a natural extension
\beq\label{Phi}
	\Phi := q(\t\Phi): {\cal V}\to {\cal V}_\Phi
\eeq
which provides the algebra ${\cal V}_\Phi$, with associated maps $q_\Phi$ and isomorphism $\ii_\Phi:\t V\otimes \t V \to {\cal V}_\Phi$ (see (\ref{iK})). We have, explicitly, $\Phi = \ii_\Phi\circ \t\Phi\circ \ii^{-1}$ (see (\ref{qK})). This means that the map $\Phi^{-1}$ intertwines both the local degrees of freedom and the local structure of the canonical algebra ${\cal V}$ in order to give the physical algebra ${\cal V}_\Phi$

Given $\t\Phi$, the physical time evolution and physical state are those obtained by extending $(\t U_t)_\Phi:=\t\Phi\circ \t U_t\circ \t\Phi^{-1}$ and $(\t\sigma_{\beta_1,\beta_2})_{\Phi}:=\t\sigma_{\beta_1,\beta_2}\circ \t\Phi^{-1}$, the latter for appropriate $\beta_1$ and $\beta_2$. Following (\ref{qKmap}), these extensions are respectively
\beq
	q_\Phi\big((\t U_t)_\Phi\big) = \Phi\circ U_t\circ \Phi^{-1}\quad\mbox{and}\quad
	q_\Phi\big((\t\sigma_{\beta_1,\beta_2})_{\Phi}\big) = \sigma_{\beta_1,\beta_2}\circ \Phi^{-1}.
\eeq

In general, $\t\Phi$ is {\em not} compatible with the canonical local structure, so that ${\cal V}_\Phi$ represents a different local structure than that on the (double) line $\R\cup_{\rm disj}\R$. If it is, then $\Phi = \ii \circ \t\Phi\circ \ii^{-1}$ is a ${\cal V}$-isomorphism, so that ${\cal V}_\Phi\cong {\cal V}$. In this case, we recover the local structure representing topology of the line.

For describing the system with a topological defect, we simply choose $\Phi = {\rm id}$, in which case $\Phi$ is a (trivial) automorphism. 

In order to describe the reflecting defect, we set $\t\Phi = \t\Phi_o$ defined by
\beq\label{tPhio}
	\t\Phi_o: \ba{rcll}
	v(x)\otimes {\bf 1}&\mapsto& v(x)\otimes {\bf 1} & (x\leq 0)\\
	v(x)\otimes {\bf 1}&\mapsto& {\bf 1}\otimes v(-x) & (x>0)\\
	{\bf 1}\otimes v(x)&\mapsto& v(-x)\otimes {\bf 1} & (x<0)\\
	{\bf 1}\otimes v(x)&\mapsto& {\bf 1}\otimes v(x) & (x\geq 0)
	\ea
\eeq
for every $v\in V$, with supports transforming as
\beq
	\t\Phi_o:
	\lt(A_-\cup A_+\rt) \cup_{\rm disj} \lt(B_-\cup B_+\rt) \mapsto \lt(A_-\cup (-B_-)\rt) \cup_{\rm disj} \lt(
	(-A_+)\cup B_+\rt)
\eeq
where $A_- \subset (-\infty,0]$, $A_+\subset (0,\infty)$, $B_-\subset(-\infty,0)$ and $B_+\subset [0,\infty)$. This has, physically, the effect of ``unfolding'' the left-hand side into right-moving currents, and the right-hand sides into left-moving currents (or, of folding the right-moving line into the left-hand side, and the left-moving line into the right-hand side). Writing $\tV \otimes \tV = (\tV_{\leq 0}\otimes \tV_{>0}) \otimes (\tV_{<0}\otimes \tV_{\geq 0})$ in an obvious notation, the $\tV\otimes \tV$-automorphism $\t\Phi_o$ is essentially a permutation of the middle two tensorial factors (with appropriate permutation of the associated supports). Note that there are other consistent choices for the inclusion of the boundary point $x=0$, and that this is largely irrelevant for our discussion.

The full local field algebra in the case of a reflecting defect is then ${\cal V}_o := {\cal V}_{\Phi_o}$ and the corresponding physical intertwiner
\beq\label{Phio}
	\Phi_o := q(\t\Phi_o) : {\cal V}\to{\cal V}_o.
\eeq
We note that $\t\Phi_o$ is {\em not} compatible with the canonical local structure. This is physically because $\h V\otimes \h V$ contains the OPEs representing the topology of the line, in particular there is a nontrivial OPE expanding $h_+(-x) h_+(x)$ for $x>0$ into a graded series of fields at $x$; while with a cut, we do not expect $h_+(-x)$ and $h_+(x)$ to interact in any way.

The time evolution on ${\cal V}_o$ compatible with the reflecting defect is then
\beq\label{uto}
	U_t^o := \Phi_o\circ U_t\circ \Phi_o^{-1}.
\eeq
Thanks to  the unfolding performed by $\Phi_o$, according to this time evolution there is reflection on the boundary at $x=0$ both for fields coming from the left and fields coming from the right. Although $U_t^o$ is not a time evolution where $v(x)\otimes {\bf 1}$ is always right-moving and ${\bf 1}\otimes v(x)$ is always left-moving, still for every $x\neq 0$ there is a $t'>0$ such that for every $v,w\in V$ and $|t|<t'$, we have $U_t^o(v(x)\otimes w(x)) = U_t(v(x)\otimes w(x)) = v(x-t)\otimes w(x+t)$. This can be generalized to other elements as long as they are supported away from the point $x=0$, so that $U_t^o$ is still, locally and away from the defect, a time evolution where the first factor is right-moving and the second left-moving. The only difference with $U_t$ occurs at the point $x=0$. In terms of our heuristic discussion, this time evolution is obtained from the hamiltonian $H_o$, so that $U_t^o(\cdots) \leftrightarrow e^{iH_o t}\cdots e^{-iH_o t}$. 

Finally, the unfolding also guarantees that the state $\omega_o$ where left and right subsystems have been separately thermalized can be represented by
\beq\label{omegao}
	\omega_o = \sigma_{\beta_l,\beta_r}\circ\Phi_o^{-1}.
\eeq
By construction, this state is preserved by the time evolution $U_t^o$, $\omega_o\circ U_t^o = \omega_o$. It is worth noting that $\omega_o$ factorizes on the left- and right-subalgebras, thanks to the exchange of the middle tensor factors produced by $\Phi_o$ as described above, and to the factorization of $\sigma_{\beta_1,\beta_2}$ on the first and second tensor factors of $\h V\otimes \h V$.


\section{$S$-matrix and the non-equilibrium steady state}\label{sectness}

\subsection{The CFT non-equilibrium steady state}

As explained in the heuristic discussion, the real-time (hamiltonian-reservoir) construction of a non-equilibrium steady state with thermal flow through a topological defect is obtained by starting with two systems with boundaries, thermalized at different temperatures, which are then connected by a topological defect and evolved for an infinite time according to the time evolution of a system with a topological defect. The resulting steady state is a state with energy flow on the line, and is stationary with respect to the evolution on the line $U_t$.

In the present subsection, we show that this large-time limit exists weakly on ${\cal V}$, we establish the form of the resulting stationary state $\omega_{\rm stat}$, stationary with respect to $U_t$, and we show that it can be obtained from the initial state $\omega_o$ by composing with a ``scattering matrix'' (or scattering map) ${\cal S}:{\cal V}\to{\cal V}_o$ which is an algebra isomorphism (this corresponds, in our heuristic discussion, to ${\rm Ad}(S^{-1})$, whose action is given by (\ref{SonO})). In particular, the scattering matrix is {\em not} an automorphism of the initial algebra $ {\cal V}_o$, which indicates that it implements a change of the local structure of the algebra of observables. It is an automorphism of an algebra formed by a common subspace of ${\cal V}$ and ${\cal V}_o$, denoted $\h V_0\otimes \h V_0$, where essentially the point $x=0$ has been taken out altogether. But this algebra has a ``weaker'' local structure which is not preserved by the time evolution $U_t$. The fact that the space topology corresponding to the local structure preserved by the scattering matrix is weaker than that of both defects is an indication that certain QFT renormalization processes are modified in the steady-state limit. This reflects the fact that certain high-energy states may occur after the steady-state limit which are not described by CFT, and that the associated observables do not have a steady-state limit in the CFT context. We will discuss this, and other, aspects in more detail in the next subsection.

In order to describe the algebra $\h V_0\otimes \h V_0$ and in fact to discuss the long-time limit generating the steady state, we now make the following construction. For every $s\in \R$ consider the subalgebra $\h V_{<s}$ of $\h V$ formed by all elements supported in (i.e.~possessing supports in) $(-\infty,s)$, and the subalgebra $\h V_{>s}$ of $\h V$ formed by all elements supported in $(s,\infty)$. The algebra $\h V_s:=\h V_{<s}\otimes \h V_{>s}$ has a local structure representing a topology where a single point has been taken out at $x=s$, thus separating the left- from the right-hand side of $s$; OPEs are implemented just for local fields all standing on the left or all standing on the right of the point $x=s$. Note that this is not like a reflecting defect, because we have only one copy of the local algebra, so there is not reflection phenomenon. The state $\sigma_\beta$ defined in (\ref{defsigma}) gives rise to a well-defined linear functional on $\h V_s$ for every $s$. However, the time-evolution operators $U_{t}^\vee$ as defined in (\ref{defut}) are not automorphisms of $\h V_s$. This is because right-movers will eventually go through the point $x=s$. Yet $U_t^\vee$ provide isomorphisms between the algebras $\h V_s$, and we have
\beq\label{UtVs}
	U_{t}^\vee:\h V_s\to \h V_{s-t},\quad s,t\in\R.
\eeq

Now form the space $\h V_s\otimes \h V_{-s}$. This is a sub-algebraic structure for ${\cal V}$. Hence the state $\sigma_{\beta_1,\beta_2}$ is a linear functional on it. However, the physical state representing the reflecting defect $\omega_o$ in which we are interested is generically {\em not}. In fact, it is a linear functional on $\h V_s\otimes \h V_{-s}$ if and only if $s=0$. This is because for $s=0$, there is no OPE implemented between fields on the left and on the right of the defect, so $\omega_o$ is well defined on the algebra. 

We find that $\h V_0\otimes \h V_0$ is a sub-algebraic structure for both ${\cal V}$ and ${\cal V}_o$, and that $\Phi_o$ as defined in (\ref{Phio}) is a $\h V_0\otimes \h V_0$-automorphism:
\beq\label{v0v0}
	\Phi_o: \h V_0\otimes \h V_0 \to \h V_0\otimes \h V_0. 
\eeq
We note also that the limit $\lim_{s\to\infty}\h V_s\otimes \h V_{-s}$ exists in a local sense and gives ${\cal V}$ (in particular, for every $\zeta\in {\cal V}$ there exists a $s'$ such that $\zeta\in \h V_s\otimes \h V_{-s}$ for all $s>s'$).

We can now express our main proposition.
\begin{propo}\label{propolimit}
The limit
\beq\label{limit}
	\lim_{t\to\infty} \omega_o\big[U_t(\zeta)\big] = \omega_{\rm stat}\big[\zeta\big]
\eeq
exists and is finite for every $\zeta\in{\cal V}$, and gives rise to a state $\omega_{\rm stat}$ on ${\cal V}$ given by
\beq\label{omegastat}
	\omega_{\rm stat} = \sigma_{\beta_l,\beta_r}.
\eeq
\end{propo}
\proof
First, we observe that thanks to (\ref{UtVs}) and (\ref{ut}), the state $\omega_o\circ U_t$ is well defined on $V_t\otimes V_{-t}$. Hence by the discussion above, if the limit exists, it provides a state on ${\cal V}$. Further, since $\omega_o$ is invariant under $U_t^o$, then we may evaluate
\[
	\lim_{t\to\infty} \omega_o\big[\lt((U_t^o)^{-1}\circ U_t\rt)(\zeta)\big].
\]
It is sufficient to consider $\zeta = \ii(\t\zeta)$ for every $\t\zeta\in\tV\otimes\tV$, and it is sufficient to take generating elements in $\tV\otimes \tV$. We now evaluate the action of $(U_t^o)^{-1}\circ U_t$ on generating elements $v(x)\otimes{\bf 1}$ and ${\bf 1}\otimes v(x)$, for $v\in V$ and $x\in\R$. Thanks to (\ref{ut}), for every $x$, there exists a $t$ large enough such that $U_t(v(x)\otimes {\bf 1})$ is supported on $(\infty,0)$ and $U_t({\bf 1}\otimes v(x))$ is supported on $(0,\infty)$. Then, using (\ref{tPhio}) and (\ref{Phio}), we see that on these vectors $\Phi_o \cong {\rm id}$. Hence, using (\ref{uto}) we find that for every $x$ there exists a $t$ large enough such that $\lt((U_t^o)^{-1}\circ U_t\rt)(v(x)\otimes {\bf 1}) = \lt(\Phi_o \circ U_t^{-1}\circ U_t\rt)(v(x)\otimes {\bf 1}) = \Phi_o(v(x)\otimes {\bf 1})$ and similarly $\lt((U_t^o)^{-1}\circ U_t\rt)( {\bf 1}\otimes v(x)) = \Phi_o({\bf 1}\otimes v(x))$. Since this holds for all generating elements, this shows that
\[
	\lim_{t\to\infty} \omega_o\big[U_t(\zeta)\big] = \lt(\omega_o\circ \Phi_o\rt)\big[\zeta\big].
\]
Using (\ref{omegao}), this completes the proof.
\eproof

This can be re-expressed in terms of the scattering matrix ${\cal S}$.
\begin{propo}\label{propoS}
We have that
\beq
	{\cal S}:=\lim_{t\to\infty} \lt((U_t^o)^{-1}\circ U_t\rt) = \Phi_o:{\cal V} \to {\cal V}_o
\eeq
in the sense that for every $\zeta\in{\cal V}$, there is a $t$ large enough such that the left-hand side of the following is well defined and
\beq
	\lt((U_t^o)^{-1}\circ U_t\rt)(\zeta) = \Phi_o(\zeta).
\eeq
Then,
\beq
	\omega_{\rm stat} = \omega_o\circ {\cal S}.
\eeq
\end{propo}
\proof This follows immediately from the discussion above and the proof of Proposition \ref{propolimit}.
\eproof

We finally note that thanks to (\ref{v0v0}) and the discussion around it, we have that ${\cal S}$ is a $\h V_0\otimes \h V_0$-automorphism:
\beq
	{\cal S}: \h V_0\otimes \h V_0 \to \h V_0\otimes \h V_0.
\eeq
That is ${\cal S}$ preserves a weaker local structure than that of the full algebras ${\cal V}$ or ${\cal V}_o$.

\subsection{Universality and scaling limit}\label{ssectuni}

It is common belief, without any reasonable doubt but in most cases without formal proof, that 1+1-dimensional CFT describes the low-energy sector of quantum chains (one-dimensional chain of quantum degrees of freedom connected by few-neighbour interactions) at critical points with dynamical exponent $z=1$. That is, in such critical systems and in the thermodynamic limit, whenever only states of energies much smaller than the typical microscopic interaction energies (or, in conduction electron systems, the bandwidth) $D$ are excited, then the resulting physics is correctly described by CFT. This includes the low-temperature equilibrium thermodynamics as well as the large-distance coordinate dependence of equilibrium correlation functions of local observables. The description provided by a given CFT model is common to every microscopic model in a universality class: the low-energy sector is universal. 

The situation is less clear concerning out-of-equilibrium problems. An early comment was that dangerous irrelevant operators may play a role and modify the low energy description \cite{Giama92}. But another subtlety in the real-time hamiltonian-reservoir situation is the question as to the existence of a nontrivial (current-carrying) steady state. Indeed, in this construction, the effective baths are, after an infinite time, asymptotically far. If transport is purely diffusive, then Fourier's law apply and the zero gradient of temperature implies that there is no energy current. However, there is up to now no consensus in the literature whether transport in 1D quantum system, especially in the low-temperature regime, is predominantly ballistic or diffusive. Indeed, some time ago ref.\cite{Cast95} claimed that transport in integrable 1D systems is ballistic and that this is a generic and characteristic property of integrable systems. Since CFT is integrable (in fact, momentum conservation alone should be sufficient), if the low-temperature regime is described by CFT out of equilibrium, this would imply a non-negligible ballistic component even for non-integrable microscopic chains. However, later ref.\cite{Aff05} argued that diffusion is universal, even for integrable systems, and related this property to the existence of a large relaxation time. More recently, ref.\cite{Pros11} obtained an exact bound on the so-called Drude weight in integrable models of spin chains driven by Lindblad operators, which implies a ballistic transport. This was numerically confirmed in \cite{Moore12}. But non-zero Drude weights in non-integrable systems were also found in \cite{Moore12}, implying ballistic transport for such systems as well.

This latter surprising fact may indeed suggest that CFT describes the low-temperature regime, its integrability providing a ballistic channel for energy transport. However, in actual quantum chains, the scaling limit is never {\em exact}, hence there may be a large relaxation time $T_{\rm relax}$ depending on microscopic data (the band curvature for instance) beyond which transport becomes diffusive. That is, we may expect transport, at low temperatures, to be essentially ballistic at times large but smaller than $T_{\rm relax}$, and to become diffusive at times larger than $T_{\rm relax}$ for non-integrable microscopic systems. The ballistic behaviours seen in \cite{Moore12} for non-integrable systems would then be due to the fact that the simulation times were long enough to reach quasi-stationary states but smaller than the relaxation time $T_{\rm relax}$ above which the systems would have behaved diffusively.

We now discuss how this intriguing, but complicated, matter may be analysed in our set-up. Ultimately, we would like to prove (under natural assumptions) that the CFT description of the non-equilibrium steady state we have given above correctly and exactly describes non-equilibrium steady states of critical quantum chains (as obtained in a real-time construction with hamiltonian reservoirs) occurring in the limit of low temperatures $T_{l,r}$ and large times. Our CFT results are, by construction, obtained in the low energy limit followed by the large time limit. But more precisely, we would like to show that, as the energy is lowered, there is a region of time which becomes larger and larger and is essentially limited by $T_{\rm relax}$, in which the averages in the quantum chain are more and more accurately described by the CFT results. If $T_{\rm relax}$ is infinite (for instance, if the quantum chain is integrable), then we would expect that we can in fact exchange the low energy and large time limits. The non-commutation of the limits would signal the existence of the large relaxation time $T_{\rm relax}$, in which case our CFT description would be, at small but non-zero temperatures, approximately valid out of equilibrium only at intermediate time scales\footnote{These time scales are nevertheless relevant both in numerical simulations and in experiments as $T_{\rm relax}$ may be very large.} large enough to reach a quasi-stationnary state but smaller than $T_{\rm relax}$. The aim of the following discussion is to present steps towards a possible proof that the limits can be interchanged, in order to pinpoint where the theoretical difficulties are located and thus help in understanding how a time scale $T_{\rm relax}$ may appear.

Given an observable in the quantum chain, we make three assumptions. The first assumption is that the stationary limit exists for every temperature $T_l$ and $T_r$. This, thanks to the Gell-Mann-Low theorem, implies that when $T_l=T_r$, the limit gives the equilibrium Gibbs state of the connected system. The second assumption is that the low-energy sector is described by CFT, in agreement with common belief. The third assumption is that with $T_l=T_r$, the low-temperature limit of the equilibrium average of the observable, both before contact and after the stationary limit has been reached, converges to the CFT average. 
From this, we argue that our CFT computations describe correctly the low-temperature limit, $T_{l,r}\ll D$, of the non-equilibrium average provided an exchange of limit specified below is valid. Note that the low-temperature scaling limit necessitates space rescaling and in general a renormalization of the observable -- we will understand the observable as flowing with the scale of the temperatures in an appropriate way, and discuss the subtleties related to this.

To get a hint on the usual argument for the convergence toward CFT at low energy, let us write the equilibrium quantum chain Gibbs state at temperature $T=\beta^{-1}$ as $\omega_o^{\rm chain}:=\mathfrak{z}^{-1}_\beta\,\int d\nu_o^{\rm chain}(E)\, e^{-\beta E}\, \omega^{\rm chain}_{o|E}$, where $\omega^{\rm chain}_{o|E}$ is the normalised state that traces over the subspace of energy $E$, $d\nu_o^{\rm chain}(E)$ is the energy density in the thermodynamic limit, and $\mathfrak{z}_\beta$ the normalisation factor. Let $\Or_{\rm chain}$ be an observable on the chain, whose average we wish to evaluate. Provided that the energy density does not grow too fast at high energy $E\gg D$, and likewise that the trace $\omega^{\rm chain}_{o|E}\big[\Or_{\rm chain}\big]$ does not grow too fast, all contributions from energies $E\gg D$ are exponentially suppressed by the factor $e^{-\beta D}$ in the expectation $\omega^{\rm chain}_o\big[{\cal O}_{\rm chain}\big]$. Then, with an appropriate choice of $\Or_{\rm chain}$, renormalized appropriately with the scale $\beta D$, the usual assumption is that $\omega^{\rm chain}_o\big[\Or_{\rm chain}\big]$ converges towards an average in the CFT Gibbs state $\omega_o$ as $\beta D\to\infty$ of an appropriate CFT field (or product thereof) $\Or$ corresponding to $\Or_{\rm chain}$.

The hamiltonian-reservoir real-time description of the non-equilibrium steady state we presented above may be applied to the quantum chain. The initial state $\omega_o^{\rm chain}$ represents a direct product of Gibbs states at temperatures $T_l$ and $T_r$ for two separate half-infinite critical quantum chains, with hamiltonian $H_o^{\rm chain}=H_l^{\rm chain}+H_r^{\rm chain}$, and the evolution hamiltonian is $H^{\rm chain}$, where the interaction links between left and right half-chains have been added in order to make a homogeneous infinite chain. We may represent $\omega_o^{\rm chain}$ as above, but separating the energies of each half
\beq\label{omcha0}
	\omega_o^{\rm chain} =\mathfrak{z}^{-1}_{\beta_1,\beta_2}\,\int d\nu_o^{\rm chain}(E_l,E_r)\, e^{-\beta_l E_l-\beta_r E_r} \omega^{\rm chain}_{o|E_l,E_r},
\eeq
where $\omega^{\rm chain}_{o|E_l,E_r}$ corresponds to a trace over the $H_l^{\rm cha},H_r^{\rm cha}$-eigensubspace with fixed $E_l,E_r$. 

Then, the quantum chain steady state $\omega^{\rm cha}$ may be written as
\beq\label{ochastat}
	\omega^{\rm chain}_{\rm stat}\big[{\cal O}_{\rm chain}\big] = \omega^{\rm chain}_o\big[S_{\rm chain}^{-1} {\cal O}_{\rm chain} S_{\rm chain}\big]
\eeq
for an appropriate family of operators ${\cal O}_{\rm chain}$ (e.g.~operators supported on a finite number of sites), where $S_{\rm chain} = \lim_{t\to\infty} e^{-itH^{\rm chain}} e^{itH_o^{\rm chain}}$ is the lattice $S$-matrix. We assume that the stationary limit $t\to\infty$ exists for all $\beta_l,\beta_r$. This implies that the stationary limit exists in $\omega^{\rm chain}_{o|E_l,E_r}\big[S_{\rm chain}^{-1} {\cal O}_{\rm chain} S_{\rm chain}\big]$ independently for every $E_l,E_r$. From Gell-Mann-Low's theorem, the $S$-matrix intertwines the hamiltonians $H_o$ and $H$ \footnote{We are thus implicitly assuming that there is no bound states localised around the junction after the sub-systems have been connected. This is reasonable because, if such bound states appear they should nevertheless not contribute in the scaling limit probing energy smaller than the binding energy.},
\beq \label{Sinter}
H^{\rm chain}S_{\rm chain} = S_{\rm chain}H_o^{\rm chain}.
\eeq
This implies that $S_{\rm chain}$ maps each $H_l^{\rm cha},H_r^{\rm chain}$-eigenstate of eigenvalue $E_l,E_r$ to an $H$-eigenstate of eigenvalue $E_l+E_r$. Therefore, for every $E_l,E_r$, there is a subspace of the $H$-eigenspace of energy $E=E_l+E_r$, such that the associated density is
\[
	d\nu^{\rm chain}(E_l,E_r) = d\nu_o^{\rm chain}(E_l,E_r),
\]
and such that the associated trace is
\[
	\omega^{\rm chain}_{E_l,E_r} =
	\omega^{\rm chain}_{o|E_l,E_r} \circ {\rm Ad}\lt(S_{\rm chain}^{-1}\rt).
\]
Hence,
\beq\label{omcha}
	\omega^{\rm chain}_{\rm stat} = \mathfrak{z}^{-1}_{\beta_1,\beta_2}\,\int d\nu^{\rm chain}(E_l,E_r)\, e^{-\beta_l E_l-\beta_r E_r} \omega^{\rm chain}_{E_l,E_r}.
\eeq
With $\beta_l=\beta_r=\beta$, we see that $\omega^{\rm chain}_{\rm stat}$ is the Gibbs state for the hamiltonian $H$ at inverse temperature $\beta$.

We now consider the low-temperature limit. The assumption that the low-temperature limit exists and is described by CFT at $\beta_l=\beta_r$ both before the contact (i.e.~with $\omega^{\rm chain}_o$) and in the stationary limit (with $\omega^{\rm chain}_{\rm stat}$), and the assumption that the low-energy sector is described by CFT, imply that in the integrals (\ref{omcha0}) and (\ref{omcha}), the low-temperature limit is obtained from the low-energy sector $E_l+E_r\ll D$. This means that at large energies, the averages under $\omega^{\rm chain}_{o|E_l,E_r}$ and $\omega^{\rm chain}_{o|E_l,E_r}$ and the energy density do not overcome the exponential suppression, and that the renormalization is that necessary for the limit to be finite. Hence this also holds with $\beta_l\neq \beta_r$: as both temperatures are lowered, only the low-energy sector is involved, so the result is described by CFT. It is clear that our CFT construction for $\omega_o$ describes the low-temperatures limit of $\omega^{\rm chain}_o$. 
In order to argue that the low-temperatures limit of $\omega^{\rm chain}_{\rm stat}$ is $\omega_{\rm stat}$, we need to make sure that the limits of large time and of low temperatures can be interchanged, because we need to be able to argue that the conformal $S$-matrix constructed in previous sections faithfully represent the action of $S_{\rm chain}$ in $\omega^{\rm chain}_{o|E_l,E_r}$ at low energy.
Naively, since both at the beginning and at the end of the process, only low-energy states are necessary to describe the average when temperatures are low, this suggests that throughout the process  the low-energy sector is sufficient to describe the low-temperature limit. Here, however, more complete arguments, which we do not presently have, would be needed.

The arguments above make it clear that $\omega_{\rm stat}$ can only describe the low-temperature scaling limit of the quantum chain steady state for observables whose renormalization is the same before the contact and in the stationary state; otherwise we cannot guarantee that the process from the former state to the latter state corresponds to an adiabatic change, hence we cannot interchange the low-temperatures and large-time limit. 
Due to the local quench, local operators near the origin will have different renormalizations. For instance, the product of degrees of freedom on the left-hand side by degrees of freedom on the right-hand side, both near the origin, renormalizes differently before contact than in the stationary state. This means that we have to restrict the set of quantum chain observables for which our results describe the correct low-temperatures limit. This restriction is the quantum-chain underlying phenomenon corresponding to the reduction of the algebra of CFT observables from ${\cal V}$ to $\h V_0\otimes \h V_0$.
 
\subsection{Approach to the steady state}
To decipher how the system state converges towards its steady states we have to look at the large but finite time $t_o$ behavior of the correlation functions 
\[\omega_o\big[\prod_j h_+(x_j,t_o) \prod_k h_-(y_k,t_o) \big],\]
where the time evolution is defined by the $H$-dynamics. 
As soon as $t_o$ is larger than the maximum of the distances of the field locations from the contact point, that is $t_o>{\rm max}[|x_j|,\,|y_k|]$, all the chiral operators $h_\pm$ have moved either to the left or the right subsystem since then $x_j-t_o<0$ for all $j$ and $y_k+t_o>0$ for all $k$. The above correlation functions then factorize on left and right sectors as
\[\omega_o\big[\prod_j h_+(x_j-t_o)\big]\cdot\omega_o\big[ \prod_k h_-(y_k+t_o) \big],
\quad t_o>{\rm max}[|x_j|,\,|y_k|],\]
and these two factors are $t_o$ independent due to the $H_o$-invariance of the $\omega_o$ measure (recall that translating $h_\pm$ in the $\omega_o$-correlation functions using the $H_o$-dynamics may require using the reflecting boundary conditions $h_\pm(x)=h_\pm(-x)$).

Hence, on hamiltonian and momentum densities (and their descendants) the approach to the steady state is abrupt. It is instantaneous as soon as the observable finds itself in the steady state, which occurs when the sharp left or right boundary of the steady state region, traveling at the speed of light (Fermi velocity) from the origin, has crossed the observable's position. This takes a time equal to the distance of the observable to the origin (with unit speed of light). This is valid in the scaling limit. In the quantum chain model with a microscopic energy scale $D$, we expect two characteristic time scales, the large infrared time scale of the order of the size of the observation domain associated to an abrupt change as we just described, and a short non-universal time scale of order $1/D$ associated to exponential convergences, as seen numerically \cite{Moore12}. Note that these time scales are not (or perhaps very weakly) temperature dependent, contrary to the relaxation time scale $T_{\rm relax}$ discussed in the previous Subsection, which is expected to grow as the temperature scale is lowered.

\section{Energy flow and its large deviation function}\label{sectenergy}

We now evaluate physical quantities (the energy current and the large-time statistics of the fluctuations of energy transfer) using the nature of the stationary state we constructed above. 

\subsection{The energy current}\label{ssectcurrent}

By standard field theory arguments, the local operator whose average gives the energy current is the momentum density operator (\ref{momdens}),
\beq
	J_E(x)= p(x) = h_+(x) - h_-(x)
\eeq
with $h_+(x)$ and $h_-(x)$ the right-moving and left-moving chiral energy densities (\ref{hphm}). Thanks to Proposition \ref{propolimit}, the stationary average is $\omega_{\rm stat}\big[J_E(x)\big] = \sigma_{\beta_l,\beta_r}\big[J_E(x)\big]$, and thanks to (\ref{hpmxt}) and (\ref{sym}), it is independent of $x$, so we can choose $x=0$. Then, thanks to the factorization (\ref{sigma2}), and using the expression (\ref{cfT1}), (\ref{defsigma}) as well as translation invariance on the plane (i.e.~the $L_{-1}$-derivative property of vertex operator algebras and the fact that $L_{-1}{\bf 1}=0$), we have
\beq
	\omega_{\rm stat}\big[J_E(0)\big] =
	\big({\bf 1},Y(\frak{h}^{(\beta_l)},0){\bf 1}\big) -
	\big({\bf 1},Y(\frak{h}^{(\beta_r)},0){\bf 1}\big).
\eeq
Finally, using (\ref{transfoh}) and the fact that $({\bf 1},Y(\frak{h},0){\bf 1})=0$, we find
\beq\label{JE0}
	\omega_{\rm stat}\big[J_E(0)\big] = \frc{\pi c}{12} \lt(\beta_l^{-2}-
	\beta_r^{-2}\rt)
\eeq
in agreement with the expression for $\bra J_E\ket_{\Delta\beta\neq0,\Delta\mu=0}$ quoted in Section \ref{sectresults}.

\subsection{Large-time energy transfer statistics}

\subsubsection{Definition, measurement protocols, proposition}

Let us refer for a moment to the heuristic discussion of Section \ref{sectheuristics}. In order to evaluate the large-time statistics of the fluctuations of energy transfer, we need to define an operator whose time derivative (with respect to the $H$-evolution) is the energy current, and evaluate the statistics of the change this operator undergoes after a large time $t$. A natural candidate is the operator that measures half the total energy difference between the left- and right-hand side:
\beq\label{E}
	E = \frc12 \lt(H_o^r - H_o^l\rt),\quad \frc{d E(t)}{dt} = J_E(0,t)=h_+(-t)-h_-(t).
\eeq
After the infinite-length limit $R\to\infty$ has been taken, the operator $E$ does not make sense anymore, as it then measures the difference between infinite quantities; technically, it also falls outside our formalism, as it is not a local operator, and in particular it does not have a stationary limit. Yet, after a finite time $t$, its change is
\beq\label{EtE}
	\Delta E(t):=E(t)-E=\int_{0^+}^t ds\, J_E(0,s) = \int_{0^+}^t dx\,(h_+(-x)-h_-(x))
\eeq
 which is expressed in terms of local fields and which possesses an infinite-length and stationary limit. This suggests that the statistics of the change it undergoes after a finite time $t$ can be evaluated in the non-equilibrium stationary state using our local-field formalism. We now explain how this is done.

In order for the initial part of the derivation to make sense, let us first consider a finite-length quantum chain (hence in particular before taking the scaling limit, in order to avoid UV divergences), on which we have an operator $E$ defined as in (\ref{E}). One has to be careful as to how to define the energy transfer during a time $t$. An experimentally relevant and standard way is to use an {\em indirect} measurement protocol: essentially, the current $J_E$ is linearly coupled to another quantum system, a ``dial'' that increases proportionally with time with a proportionality constant given by $J_E$. Quantum measurements are performed on this dial, and from this, the statistics of the energy transfer is inferred. In particular, one infers the probability $P_t(\Delta e)$ that the energy transferred after a time $t$ be $\Delta E(t) = \Delta e$, and defines
\[
	{\cal Z}_t(\lambda):= \sum_{\Delta e}e^{i\lambda \Delta e}\, P_t(\Delta e) 
\]
where $\lambda$ is a {\em formal variable} (so that this is the generating function of averages of powers of $\Delta e$). One then constructs the large deviation function 
\beq\label{Find}
	F_{\rm ind}(\lambda) := \lim_{t\to\infty} t^{-1} \log{\cal Z}_t.(\lambda)
\eeq
This indirect-measurement setup was considered in \cite{LLformula} in the case of charge transfers, where more explanations can be found. Adapting the expression found in \cite{LLformula} to the case of energy transfer, we then expect
\beq \label{Finddef}
F_{\rm ind}(\lambda) = \lim_{t\to\infty} t^{-1}\log \lt(
	\Tr\lt(\rho_{\rm stat}\, e^{i\lambda E(t)}e^{-i\lambda E}\rt)\rt).
\eeq

In fact a formula very similar to (\ref{Finddef}) can be obtained if we consider standard von Neuman measurements of $E$, instead of indirect measurements, as follows. Assume two measurements of $E$ are performed: First $E$ is measured exactly at the original contact time $-t_o$. The output is $e_1$ with probability ${\rm Tr}(P_{e_1}\rho)$, where $P_{e_1}$ is the projector on the corresponding eigenspace. Then, after a long time, at time $0$, $E$ is again measured. The output is $e_2$ with probability $P_{t_o}(e_2,e_1)={\rm Tr}(P_{e_2}e^{-it_oH} P_{e_1}\rho_o P_{e_1} e^{it_oH}P_{e_2})$. Since $\rho_o$ commutes with $P_{e_1}$, using the main property of projectors we can simplify this to ${\rm Tr}(P_{e_2}e^{-it_oH} P_{e_1}\rho_o  e^{it_oH})$, which we transform into ${\rm Tr}(P_{e_2}(t_o) P_{e_1}\rho_o)$. Using $\sum_e f(e) P_e = f(E)$ and $P_t(\Delta e) = \sum_{e_1,e_2:e_2-e_1=\Delta e} P_t(e_2,e_1)$, we obtain
\beq\label{FvNdef}
	F_{\rm vN}(\lambda) = \lim_{t\to\infty} t^{-1}\log\lt(
	\Tr\lt(\rho_o\, e^{i\lambda E(t)} e^{-i\lambda E}\rt)\rt).
\eeq

We remark that although the operator averaged in eqs.~(\ref{Finddef})  and (\ref{FvNdef}) appears non-local, the Baker-Campbell-Hausdorff (BCH) formula guarantees that only $\Delta E(t) = E(t)-E$ and its commutators with itself and with $E$ are involved. Since the time-derivative of $E(t)$ only involves an operator supported on a finite number of sites near the origin, then $\Delta E(t)$ is essentially localized near the origin, on a region of length proportional to $t$; whence also all commutators with itself and with $E$ are. This suggests that the infinite-length and, as is needed in (\ref{Finddef}), stationary limit exists; and, in accordance with our arguments of Subsection \ref{ssectuni}, that in the scaling (low-temperatures) limit it is given by evaluating the corresponding average in the stationary state $\omega_{\rm stat}$ in CFT. In the CFT, by the BCH formula, the operator averaged in (\ref{Finddef}) and (\ref{FvNdef}) is explicitly a series of integrals of local fields supported on a region near to $[-t,t]$; whence indeed its stationary limit exists in the CFT construction itself. It is this integral expression, written in (\ref{init}), that we will use as our starting point for our actual proof of Proposition \ref{propoFCS} below.

We will show the following.
\begin{propo}\label{propoFCS}
The energy cumulant generating function (\ref{init}) in the non-equilibrium steady state $\omega_{\rm stat}$ of CFT is
\begin{eqnarray} \label{energyFCS}
F(\lambda):=F_{\rm vN}(\lambda)= F_{\rm ind}(\lambda) = \frac{c\pi}{12\hbar}\, \Big( \frac{i\lambda}{\beta_l(\beta_l-i\lambda)}-\frac{i\lambda}{\beta_r(\beta_r+i\lambda)}\Big)
\end{eqnarray}
where $\beta_{l,r}= (k_B T_{l,r})^{-1}$.
\end{propo}
Here we have put back Planck's and Boltzmann's constant. Note that the result is very universal, depending only on the CFT central charge and universal constants. The order $O(i\lambda)$ gives the current $\omega_{\rm stat}\big[J_E(0)\big] = \bra J_E\ket_{\Delta\beta\neq0,\Delta\mu=0}$ quoted in Section \ref{sectresults} and derived in (\ref{JE0}). The function (\ref{energyFCS}) satisfies the fluctuation relation \cite{CohenGal,Espo},
\begin{eqnarray}\label{fluctu}
F(i(\beta_r-\beta_l)-\lambda)=F(\lambda).
\end{eqnarray}
That the energy transport fluctuations satisfy the fluctuation relation has been checked in the quantum harmonic oscillator chain \cite{OscilChain}. As usual, the fluctuation relation relates the probabilities $P_t(\theta)$ and $P_t(-\theta)$ of opposite energy transfers $E(t)-E=\pm t\theta$ across the interface:
\[ e^{-t\beta_l\theta}\, P_t(\theta)d\theta = e^{-t\beta_r\theta}\, P_t(-\theta)d\theta.\]

We mention here that it is possible to define a two-step measurement protocol as described above, but where the first measurement time occurs when the steady state has been established (time 0), instead of at the contact time ($-t_o$). Similar arguments as those above lead to $P_t(e_2,e_1)={\rm Tr}(P_{e_2}e^{-itH} P_{e_1}\rho_{\rm stat} P_{e_1} e^{itH}P_{e_2})$. For finite lengths, $E$ has a discrete spectrum and we may use \cite{Espo} the formula $\int du \,e^{iu(E-e_1)}\propto P_{e_1}$ (with an appropriate integration range). This yields, naively, an integral representation
\beq\label{ZtZtu}
	{\cal Z}_t(\lambda)\propto \lim_{\tau\to\infty} \frc1\tau \int_{-\tau}^\tau du\, {\cal Z}_t(\lambda,u)
\eeq
with 
\begin{equation}\label{Zt}
{\cal Z}_t(\lambda,u):=\Tr\lt(\rho_{\rm stat}\, e^{-i\lt(\frc\lambda2-u\rt) E} e^{i\lambda E(t)}e^{-i\lt(\frc\lambda2+u\rt)E}\rt).
\end{equation}
Here $\tau\to\infty$ is the inverse of the typical (vanishing) energy difference between nearest states in the spectrum. Many subtleties occur in a proper argument leading to (\ref{Zt}); and further, from (\ref{Zt}), one finds that the associated large-deviation function is not (\ref{energyFCS}). See appendix \ref{appu} for calculations pertaining to this. This indicates that, due to the continuous, gapless spectrum, the two-step measurement directly in the steady state does not correspond to the same physics as the two-step measurement where the first measurement time is at the connection time. It would be interesting to understand further the physical meaning of this observation.

\subsubsection{Precise definition of the cumulant generating function}

We would like to make more precise the definition of
\beq\label{or}
	\Or =  e^{i\lambda (E+\Delta E(t))}e^{-i\lambda E}
\eeq
and of its averages as series expansions in $\lambda$, where $\Delta E(t)$ is given by \eqref{EtE}. For this purpose, we will first define a precise operator depending on a real parameter $\lambda$, whose averages will then be expanded in powers of $\lambda$ in order to obtain the cumulant generating functions.

First, in the precise definition of this operator, we require the notion of integral of local fields. The extension of the algebras introduced in Subsection \ref{topo}, like the algebra $V^\sharp$ or $\tV$, to include algebraic (formal) integral of local fields over finite intervals is completely straightforward, with in particular the supports being the (closed) integration regions. The action of the state $\sigma_{\beta_1,\beta_2}$ on expressions containing integrals is immediately interpreted using Riemann integrals, which here always exist. We will use the same notation for the algebras which include algebraic integrals.

Second, the point $0^+$ in (\ref{EtE}) corresponds to a limit process. Hence let us choose a $a>0$ small enough and define
\beq\label{DeltaEa}
	\Delta E(t,a):= \int_a^tdx\,(h_+(-x)-h_-(x)).
\eeq
It is in fact possible to show, tracing the steps below, that the large-$t$ limit commutes with the $a\to0^+$ limit; here we will simply take the limit $a\to0^+$ at the end.

Third, following the arguments in Section \ref{sect:bdry}, the heuristic definition (\ref{E}) of $E$ gives rise to the adjoint action of its exponential as
\beq\label{UtE}
	{\rm Ad}\lt(e^{itE}\rt)\mapsto U_t^E := \Phi_o\circ (U^\vee_{-t/2}\otimes U^\vee_{-t/2})\circ
	\Phi_o^{-1}.
\eeq
Note that the evolution operator $U_t^E$ preserves the initial state $\omega_o$.

Then, by a standard integral expression for the BCH formula, we may use the following expression for a regularization of $\Or$ as defined in (\ref{or}), 
\beq\label{Orexact}
	\Or\mapsto \Or_{a}(\lambda):=
	\Pexp \int_0^\lambda dz \lt(
	i\,U_z^E\lt(\Delta E(t,a)\rt)\rt).
\eeq
Here the exponential is understood as its series expansion, hence this is a graded series in the space $V^\sharp\otimes V^\sharp$ spanned by formal products of elements in $V(\R)\otimes V(\R)$ and their integrals. Here and below, the symbol ${\cal P}$ denotes the ordering according to which integrands with values of $z$ nearer to the upper limit $\lambda$ are positioned further to the right. In this expression, in every term of the graded series, $\lambda$ is a real parameter. Since we only want series expansions in $\lambda$, we will assume throughout that
\beq\label{lambda2a}
	|\lambda|<2a.
\eeq

Finally, the averages of the operator (\ref{Orexact}) will also require an appropriate UV regularization. Given an $\ep>0$ small enough, we will define regularized states $\omega_o^\ep$ and $\omega_{\rm stat}^\ep$, acting on the linear space $V^\sharp\otimes V^\sharp$, by the requirement
\beq
	\omega_{o,{\rm stat}}^\ep\big[v_0(x_0)\cdots v_n(x_n)\big]
	= \lt.\omega_{o,{\rm stat}}\big[v_0(y_0)\cdots 
	v_n(y_n)\big]\rt|_{y_k=x_k+ik\ep}
\eeq
where on the right-hand side we make an analytic continuation (it is unique and well-defined). We will define $\sigma_\beta^\ep$ similarly. This requirement corresponds to the usual Euclidean time-ordering of QFT. As a result, $\omega_{o}^\ep,\, \omega_{\rm stat}^\ep$ do not act on $\tV\otimes \tV$ because the ordering is important, but the (weak) limits $\lim_{\ep\to0^+} \omega_{o}^\ep,\, \omega_{\rm stat}^\ep$ are zero on the ideal generated by the commutators, and, seen as acting on the quotients by this ideal of the subspace where points are non-coincident in products, they give $\omega_{o}^\ep, \omega_{\rm stat}^\ep$ respectively. We will see $\Phi_o$, $U_t$, etc.~as acting on $V^\sharp\otimes V^\sharp$ in the natural way.

Hence, our starting expressions are
\beqa
	F_{\rm vN}(\lambda) &=& \lim_{\ep\to0^+} \lim_{t\to\infty} t^{-1}
	\log \lt(
	\omega_o^\ep\big[\Or_{a}(\lambda) \big]\rt)
	\n
	F_{\rm ind}(\lambda) &=& \lim_{\ep\to0^+} \lim_{t\to\infty} t^{-1}
	\log \lt(
	\omega_{\rm stat}^\ep\big[\Or_{a}(\lambda) \big]\rt).
	\label{init}
\eeqa
On the right-hand sides, at every term in the graded series $\Or_a$, the Taylor expansion in $\lambda$ is understood. It will be clear below that these Taylor expansions exist, and in fact, that the complete series in $\lambda$ obtained on the right-hand sides have a nonzero radius of convergence. We will see that the limits $t\to\infty$ are independent of $a$, and that the limits $\ep\to0^+ $ indeed exist. This latter fact indicates that the results are universal, and do not need renormalization.

\subsubsection{Heuristic proof of the Proposition} 

We first explain the lines of the proof by referring to the heuristic discussion of Section \ref{sectheuristics}. Consider the expression (\ref{stathphm}) for $\omega_{\rm stat}\big[\Or\big]$ and consider \eqref{or},
\[
	\Or =  e^{i\lambda (E+\Delta E(t))}e^{-i\lambda E}.
\]
Thanks to (\ref{EtE}), the difference $\Delta E(t)$ contains right-movers only on the left-hand side, and left-movers only on the right-hand side. By (\ref{E}) and the fact that both $H_o^r$ and $H_o^l$ act like $H_++H_-$ away from the origin and on the right-hand, respectively left-hand side, this means that all commutators involving $E$ and $\Delta E(t)$ can be obtained by replacing
$E\mapsto \frc12(H_--H_+)$. Hence, thanks to the BCH formula, we have the factorization
\beq
	\sigma_{\beta_l,\beta_r}\big[e^{i\lambda E(t)}e^{-i\lambda E}\big] = { Z}_t(\lambda;\beta_l)
	{ Z}_t(-\lambda;\beta_r)
\eeq
where
\beq\label{Ztb}
	{ Z}_t(\lambda;\beta):=
	\Tr\lt(\frak{n}\lt[e^{-\beta_l H_+}\rt]\,e^{\frc{i\lambda}2 H_+} e^{-\frc{i\lambda}2 H_++ i\lambda \int_{0}^t dx\,h_+(x)}\rt).
\eeq
The argument above also means that we may do the replacement $H_o^r\mapsto H_+$ and $H_o^l\mapsto H_-$ in the density matrix $\rho_o$ itself in (\ref{FvNdef}), which corresponds to $\rho_o\mapsto \rho_{\rm stat}$ and shows the first equation of (\ref{energyFCS}).

Define
\[
	f(\lambda;\beta) := \lim_{t\to\infty} t^{-1}\log Z_t(\lambda;\beta).
\]
Note that the argument of the limit in $t$ is to be understood as a series expansion in $\lambda$ (a generating function). Taking the $\lambda$-derivative on this series, we have
\[
	\frc{\p}{\p (i\lambda)} f(\lambda;\beta)
	=\lim_{t\to\infty} t^{-1}
	\Tr\lt(\frak{n}\lt[e^{-(\beta -i\lambda/2) H_+}e^{-\frc{i\lambda}2 H_++ i\lambda \int_0^t dx\,h_+(x)}\rt]\int_0^t dx\,h_+(x)\rt).
\]
At large times, the main contribution from the integral $\int_0^t dx\,h_+(x)$ downstairs is that of points $x$ lying far from 0 and far from $t$. These points have approximately all the same contributions, and cover a domain of length diverging like $t$. Hence we can make the replacement $\int_0^t dx\,h_+(x) \mapsto t \,h_+(t/2)$.
We can then translate in order to obtain
\[
	\frc{\p}{\p (i\lambda)} f(\lambda;\beta)
	=\lim_{t\to\infty}
	\Tr\lt(\frak{n}\lt[e^{-(\beta -i\lambda/2) H_+}e^{-\frc{i\lambda}2 H_++ i\lambda \int_{-t/2}^{t/2} dx\,h_+(x)}\rt] h_+(0)\rt).
\]
Since we are evaluating the normalized average, with a complicated density matrix, of a local operator $h_+(0)$, at large $t$, the operator $\int_{-t/2}^{t/2} dx\,h_+(x)$ in the density matrix can be replaced by its limit $\int_{-\infty}^\infty h_+(x) = H_+$. Hence, we find
\beq\label{dereq}
	\frc{\p}{\p (i\lambda)} f(\lambda;\beta)
	=\Tr\lt(\frak{n}\lt[e^{-(\beta -i\lambda) H_+}\rt]h_+(0)\rt).
\eeq
This means that
\[
	\frc{\p}{\p (i\lambda)} F(\lambda) = 
	\Tr\lt(\frak{n}\lt[e^{-(\beta_l -i\lambda) H_+}\rt]h_+(0)\rt)
	-\Tr\lt(\frak{n}\lt[e^{-(\beta_r +i\lambda) H_-}\rt]h_-(0)\rt)
	= \sigma_{\beta_l-i\lambda,\beta_r+i\lambda}\big[J_E(0)\big].
\]
Integrating the result (\ref{JE0}), we obtain (\ref{energyFCS}).

\subsubsection{Sketch of the formal proof of the Proposition}

We will use the large-distance ``cluster property'' (factorization property) of the state $\sigma_{\beta_1,\beta_2}$:
\beqa
	\lefteqn{\sigma_{\beta_1,\beta_2}\big[\zeta \xi\big]
	\to \sigma_{\beta_1,\beta_2}\big[\zeta\big]
	\sigma_{\beta_1,\beta_2}\big[\xi\big]}&& \n
	&& \mbox{if}\quad
	\exists \;A\in\Supp(\zeta),\,B\in\Supp(\xi)\;:\;
	{\rm min}(|x-y|:x\in A,y\in B)\to\infty.
	\label{factori}
\eeqa
This is a general property in QFT, and can be derived from our vertex operator algebra construction in Section \ref{sectalgebra}. We expect that it be quite a straightforward exercise to show this property, for instance, for $\zeta$ and $\xi$ products of local hamiltonian densities $h(x)$, but the details of this are beyond the scope of this paper. We also expect the corrections to factorization to be exponentially decaying for any $\beta_{1,2}<\infty$.

From (\ref{init}), we follow the lines of the heuristic proof presented above. The first step is to obtain
\beq\label{factex}
	F_{\rm vN}(\lambda) = F_{\rm ind}(\lambda) =
	f(\lambda;\beta_l) + f(-\lambda;\beta_r)
\eeq
where
\beq\label{fbl}
	f(-\lambda;\beta) := \lim_{\ep\to0^+}\lim_{t\to\infty}t^{-1}
	\log \sigma_\beta^\ep\big[
	\Pexp \int_0^\lambda dz\lt(
	-i \int_a^t dx\, \frak{h}(x+z/2)\rt)
	\big].
\eeq

Consider $F_{\rm ind}$. From the definition (\ref{tPhio}) and from (\ref{DeltaEa}), we see that for all $t>a>0$ we have $\Phi_o^{-1}(\Delta E(t,a)) = \Delta E(t,a)$, that is, $\Phi_o^{-1}$ acts as the identity. Then, we find
\beq
	\lt(U_{-z/2}^\vee \otimes U_{-z/2}^\vee\rt)(\Delta E(t,a))
	=\int_{a-z/2}^{t-z/2} dx\, h_+(-x) + \int_{a+z/2}^{t+z/2} dx\,h_-(x)
\eeq
and again $\Phi_o$ acts as the identity on this whenever $|z|<2a$, which holds here thanks to \eqref{lambda2a}. This implies that in (\ref{Orexact}) we may replace $U_z^E$ by $U_{-z/2}^\vee\otimes U_{-z/2}^\vee$. Thanks to (\ref{omegao}) (and recalling that $\omega_{\rm stat} = \sigma_{\beta_l,\beta_r}$), this also implies the first equation of (\ref{energyFCS}) (the first equality in (\ref{factex})). Using (\ref{sigma2}), we immediately obtain the second term of the second equality of (\ref{factex}). For the first term, we obtain instead
\[
	\lim_{t\to\infty}t^{-1}
	\log \sigma_{\beta_l}^\ep\big[
	\Pexp\int_0^\lambda dz \lt(
	i \int_a^t dx\, \frak{h}(-x+z/2)\rt)
	\big].
\]
We may use reflection invariance of the state $\sigma_\beta$, replacing $-x+z/2\mapsto x-z/2$, and then make the change of integration variable $z\mapsto -z$ in order to obtain $f(\lambda;\beta_l)$.

The second step is to take the $\lambda$-derivative of (\ref{fbl}), and make a translation in order to symmetrize the limits of the $x$-integrals:
\beq
	-i\frc{\p f(\lambda;\beta)}{\p \lambda}
	=\lim_{\ep\to0^+}\lim_{t\to\infty} t^{-1} \int_{-t/2+a}^{t/2} dy\,\frc{
	\sigma_{\beta}^\ep\big[ \Pexp\int_0^{-\lambda}dz\,
	\lt(-i\int_{-t/2+a}^{t/2} dx\,\frak{h}(x+z/2)\rt)\;
	 \frak{h}(y-\lambda/2)
	\big]
	}
	{\sigma_\beta^\ep\big[
	\Pexp \int_0^{-\lambda} dz\lt(
	-i \int_{-t/2+a}^{t/2} dx\, \frak{h}(x+z/2)\rt)
	\big]}
\eeq
By standard combinatoric arguments, the ratio in the $y$-integral is an infinite series of $z$- and $x$-integrals of {\em connected} averages of $z$-ordered factors:
\beqa\label{ear}
	\lefteqn{-i\frc{\p f(\lambda;\beta)}{\p \lambda}} && \\
	&=&\lim_{\ep\to0^+}\lim_{t\to\infty} t^{-1} \int_{-t/2+a}^{t/2} dy\,
	\sigma_{\beta}^\ep\big[ \Pexp\int_0^{-\lambda}dz\,
	\Big(-i\int_{-t/2+a}^{t/2} dx\,\frak{h}(x+z/2)\Big)\;
	 \frak{h}(y-\lambda/2)
	\big]_{\rm connected}.\no
\eeqa
Recall that connected averages are of the form $\sigma_{\beta}^\ep\big[h(x_1)h(x_2)\big]_{\rm connected} = \sigma_{\beta}^\ep\big[h(x_1)h(x_2)\big] - \sigma_{\beta}^\ep\big[h(x_1)\big]\sigma_{\beta}^\ep\big[h(x_2)\big]$, etc.

For convenience,  we re-write the right-hand side of (\ref{ear})  by extracting from the exponential the integration over $x$ that runs from $-\infty$ to $\infty$ (recall that the symbol ${\cal P}$ still imposes the appropriate ordering of the integrands after both exponentials are replaced by their series expression, hence this is just a notational convenience):
\beqa
	&& \lim_{\ep\to0^+}\lim_{t\to\infty} t^{-1} \int_{-t/2+a}^{t/2} dy\,\cdot\n
	&&
	\sigma_{\beta}^\ep\big[ {\cal P} \;e^{\int_0^{-\lambda}dz\,
	\lt(-i\int_{-\infty}^{\infty} dx\,\frak{h}(x+z/2)\rt)}\;
	e^{\int_0^{-\lambda}dz\,
	\lt(i\lt(
	\int_{-\infty}^{-t/2+a}+\int_{t/2}^{\infty}\rt) dx\,\frak{h}(x+z/2)\rt)}\;
	 \frak{h}(y-\lambda/2)
	\big]_{\rm connected}.\no
\eeqa
The correct definition of the average $\sigma_{\beta}^\ep\big[\cdots\Big]$ that is involved here is that where the $\pm\infty$ integration limits in the exponentials are replaced by $\pm\tau$, and where the limit $\tau\to\infty$ is taken after the average is evaluated; we keep the above notation as it is lighter.

Thanks to the cluster property (\ref{factori}), any connected average where fields are at positions that are far from each other tends to zero. Further, as mentioned, such an average tends to zero exponentially fast for any $\beta<\infty$. Hence: (i) taking the zeroth expansion term of the second exponential (i.e.~replacing it by 1), the large-$t$ limit of the second line (i.e.~for fixed $y$) exists on every expansion term of the first exponential; and (ii) the large-$t$ limit of the full integral over $y$ exists on every expansion term of the first exponential and on every expansion term of order greater than zero of the second exponential. Thanks to the factor $t^{-1}$, this implies that we may keep, in the large-$t$ limit, only the zeroth expansion term of the second exponential. By translation invariance, we may fix $y-\lambda/2$ to $0$, and this gives
\beqa
	\lefteqn{\lim_{\ep\to0^+}\lim_{t\to\infty} t^{-1} \int_{-t/2+a}^{t/2} dy\,
	\sigma_{\beta}^\ep\big[ {\cal P} \;e^{\int_0^{-\lambda}dz\,
	\lt(-i\int_{-\infty}^{\infty} dx\,\frak{h}(x+z/2)\rt)}\;
	 \frak{h}(0)
	\big]_{\rm connected}} && \n &=&
	\lim_{\ep\to0^+}
	\sigma_{\beta}^\ep\big[ {\cal P} \;e^{\int_0^{-\lambda}dz\,
	\lt(-i\int_{-\infty}^{\infty} dx\,\frak{h}(x+z/2)\rt)}\;
	 \frak{h}(0)
	\big]_{\rm connected}\n
	& =&
	\lim_{\ep\to0^+}
	\sigma_{\beta}^\ep\big[ e^{
	i\lambda \int_{-\infty}^{\infty} dx\,\frak{h}(x)}\;
	 \frak{h}(0)
	\big]_{\rm connected}.
\eeqa
Recall that since these are connected averages, and thanks to the $\ep$ regularization, every term of the average obtained by expanding the exponential exists. Every such term also has a simple dependence on $\lambda$: the $n^{\rm th}$ term is proportional to $\lambda^n$. Hence we can immediately interpret the result as a formal series expansion in $\lambda$.

We now want to evaluate $\sigma_{\beta}^\ep\big[ e^{
	i\lambda \int_{-\infty}^{\infty} dx\,\frak{h}(x)}\;
	 \frak{h}(0)
	\big]_{\rm connected}$. For this purpose, we use the expression (\ref{cfT2}), \eqref{defsigma} of the state $\sigma_\beta$ (from which the regularized state $\sigma_\beta^\ep$ is obtained), and in particular its precursor \eqref{precfT2}. Let us thus define the state $\sigma_{\beta;R}$, on $V^\sharp\otimes V^\sharp$, by attributing to $\sigma_{\beta;R}\big[v_1(x_1)\cdots v_n(x_n)\big]$ the value given by \eqref{precfT2}, and let us define the state $\sigma_{\beta;R}^\ep$ by the usual regularization. We claim that
\[
	\sigma_{\beta}^\ep\big[ e^{
	i\lambda \int_{-\infty}^{\infty} dx\,\frak{h}(x)}\;
	 \frak{h}(0)
	\big]_{\rm connected} = \lim_{R\to\infty}
	\sigma_{\beta;R}^\ep\big[e^{i\lambda
	\int_{-R/2}^{R/2} dx\,\frak{h}(x)}\;
	 \frak{h}(0)
	\big]_{\rm connected}.
\]
We omit the details of the proof of this expression, but the idea is that although clustering does not hold in finite $R$, it does hold in the simultaneous large-$R$ and large-distance limit (the cluster limit is uniform in the large-$R$ limit). Hence only the parts of the integration that are near the point 0 contribute nontrivially, and the parts that are far give vanishing corrections as $R\to\infty$.

For $\sigma_{\beta;R}^\ep$, we may use the standard algebraic construction of vertex operator algebras. We note that $\int_{-R/2}^{R/2} dx\,\frak{h}(x)$ corresponds, in the vertex operator algebra language, to the operator $(2\pi/R)(L_0-c/24)$. This operator is time-independent in any fixed time ordering. Hence, in the $\ep$-regularization and inside connected averages, we have
\[
	\sigma_{\beta;R}^\ep\big[
	e^{i\lambda \int_{-R/2}^{R/2} dx\,\frak{h}(x)}\,\frak{h}(0)
	\big]_{\rm connected} =
	\sigma_{\beta;R}^\ep\big[
	\frak{n}\lt[e^{\frc{2\pi i\lambda}{ R}L_0}\rt]
	\,\frak{h}(0)
	\big],
\]
where the normalization implied by $\frak{n}[\cdot]$ results from the definition of connected averages. The result is now independent of $\ep$, and from (\ref{precfT2}) we see that the right-hand side of the above corresponds to the shift $\beta\mapsto\beta-i\lambda$:
\[
	\sigma_{\beta;R}^\ep\big[
	\frak{n}\lt[e^{\frc{2\pi i\lambda}{ R}L_0}\rt]
	\,\frak{h}(0)
	\big]
	= \sigma_{\beta-i\lambda;R}\big[
	\frak{h}(0)
	\big].
\]
The result is then a one-point function, whose connected average is the usual average. Hence we find
\beq\label{dereqpr}
	-i\frc{\p f(\lambda;\beta)}{\p \lambda}
	= \sigma_{\beta-i\lambda}\big[ \frak{h}(0) \big].
\eeq
This is the equivalent of (\ref{dereq}), and completes the proof.

\section{Charge flow statistics and its large deviation function}\label{sectcharge}

We now extend the results of previous sections to charge transfer. This happens when the conformal chiral algebra $V$ contains an affine Kac-Moody sub-algebra on top of the Virasoro algebra~\footnote{More precisely, the chiral algebra $V$ contains the semi-direct product of the Virasoro algebra by the affine Kac-Moody algebra since the Virasoro algebra acts naturally on the Kac-Moody algebra.}. One can then pick a $u(1)$ affine sub-algebra of the Kac-Moody algebra and use it to measure what we call the charge. Generators $j_n$, $n\in\mathbb{Z}$, of the $u(1)$ affine algebra are chosen to be normalized according to $[j_n,j_m]=n\,\delta_{n+m;0}$. Let the charge number operator be the element $\frak{N}:=j_0$ of $V$. Borrowing notations from previous sections, the $u(1)$-current is identified with $\frak{j}(x) = j_{-1}{\bf 1}(x)$, normalized by the operator product expansion
\beq \label{opecurrent}
 \frak{j}(x)\, \frak{j}(y) = \frac{1}{(x-y)^2}\, {\bf 1} +\mathrm{regular}. 
 \eeq
Similarly as in previous sections, we assume having prepared the two halves of the system at equilibrium but with different temperatures and different chemical potentials (the later being coupled to the $u(1)$ charge) and glued them at a very early time.
We will construct the steady state of the total system, reached after a long waiting time, and compute the large deviation function of charge transfers. Proofs will be shorter than above since they are parallel to those used for energy transfer.

\subsection{ Heuristics and the steady state}
The physical current is actually made of a pair of chiral currents: the left and right-moving currents $j_\pm$ which according to previous settings are identified as $j_+(x):=\frak{j}(x)\otimes {\bf 1}$ and $j_-(x):= {\bf 1}\otimes\frak{j}(x)$. Their time evolutions are chiral, $(\partial_t\pm \partial_x)j_\pm=0$, so that locally $j_\pm(x,t)=j(x\mp t)$. The $u(1)$ charge that we denote by $N_{\rm tot}$ is the integral of $(j_++j_-)(x)$ along the system interval:
\[ N_{\rm tot}:= \int_{-R/2}^{0^-} dx\, (j_++j_-)(x)+ \int_{0^+}^{R/2} dx\, (j_++j_-)(x).\]
It is conserved by the time evolution if $j_+(\pm R/2)=j_-(\pm R/2)$ and if $(j_+-j_-)(x)$ is continuous at the contact point at $x=0$. The dynamics before and after contact correspond to two different choices of boundary conditions: $j_+(0^\pm)=j_-(0^\pm)$ before contact and $j_\pm(0^-)=j_\pm(0^+)$ after contact.

We start with the uncoupled density matrix $\rho_o =\rho_l\otimes \rho_r$ at time $-t_o$ with
\[ \rho_{l,r}:= \frak{n}\big[e^{-\beta_{l,r}\,(H_o^{l,r} -\mu_{l,r}\,N_o^{l,r})}\big],\]
where $H_o^{l,r}$ are the energies and $N_o^{l,r}$  the $u(1)$ charge numbers on the left/right subsystems, naturally defined by restricting the integrals of the energy densities and the currents to the subsystem intervals, and $N_{\rm tot}=N_o^l+N_o^r$. As usual, by duality this defines Gibbs states over the chiral algebras.

As above, the steady state $\rho_{\rm stat}$ is reached in the large time limit after the two subsystems have been coupled, and the $S$-matrix intertwines $\rho_o$ and $\rho_{\rm stat}$:
\[ \rho_{\rm stat}= S\, \rho_o\, S^{-1}.\]
So we need to know the action of the $S$-matrix on the $u(1)$-currents. We have:
\begin{eqnarray} \label{SonJ+}
S^{-1}\, j_+(x)\, S &=& j_+(x),~~~~~~~~~~~~~~ \ {\rm for}\ x<0,\\
S^{-1}\, j_+(x)\, S &=& j_-(-x),~~~~~~~~~~~~ \ {\rm for}\ x>0. \nonumber
\end{eqnarray}
The proof is exactly the same as for the chiral energy densities: from its definition, $S:=\lim_{t_o\to\infty} e^{-t_oH}e^{+t_oH_o}$ in the large $R$ limit, the $S$-matrix acts on operators by first evolving them with the $H$-dynamics forward in time and then backward in time with the $H_o$-dynamics, and these two only differs by the boundary condition at the contact point. Similarly,
\begin{eqnarray} \label{SonJ-}
S^{-1}\, j_-(x)\, S &=& j_+(-x),~~~~~~~~~~~~~ \ {\rm for}\ x<0,\\
S^{-1}\, j_-(x)\, S &=& j_-(x),~~~~~~~~~~~~~~~ \ {\rm for}\ x>0. \nonumber
\end{eqnarray}

Again, the left (right) movers end up in the right (left) part of the system,
and according to the usual arguments, the steady state factorizes on left/right movers,
\[ \omega_{\rm stat} = \sigma_{\beta_l,\mu_l}\otimes \sigma_{\beta_r,\mu_r} \]
where $\sigma_{\beta_{l,r},\mu_{l,r}}$ are the Gibbs states on the left/right chiral algebras at respective temperature $\beta_{l,r}$ and chemical potential $\mu_{l,r}$.

\subsection{Charge transfer}

The charge transfer that we will look at is that of half the difference of the $u(1)$ charges on the left and right subsystems, $Q := \frc12(N_o^r-N_o^l)$. In the infinite volume $Q$ is clearly infinite, but its time variation under the $H$-dynamics is finite. For $R\gg t$ we have
\[
	\Delta Q(t):=Q(t) -Q = \int_{-t}^{0^-} dx\,j_+(x)- \int_{0^+}^t dx\, j_-(x),
\]
which simply expresses the ballistic transport of charges.

To get the large deviation function of the charge transfer we must evaluate the generating function ${\cal Z}_t(\nu ) := \omega_{\rm stat}\big[ e^{i\nu (Q+\Delta Q(t))} e^{-i\nu Q}\big]$, which implicitly depends on $\beta_{l,r}$ and $\mu_{l,r}$. Following the by-now usual reasoning, ${\cal Z}_t(\nu)$ factorizes into the product of two factors associated to both chiral sectors, and we may write
\[
{\cal Z}_t(\nu) = \sigma_{\beta_l,\mu_l}\big[\Or_+\big]\, \sigma_{\beta_r,\mu_r}\big[\Or_-\big], \quad {\rm with}\
\Or_{\pm} = e^{\mp\frc{i}2 \nu N_\pm \pm i\nu\int_0^{t}dx\, j_\pm(x)}\, e^{\pm\frc{i}2 \nu N_\pm}
\]
Here $N_\pm$ refers to the charge number operators in the corresponding chiral sector (they arise because $Q_o^{r,l}$ act on left/right movers away from the origin as $N_++N_-$), and we used translation invariance to replace the integral $\int_{-t}^0$ by $\int_0^t$. 
Hence we have factorization of ${\cal Z}_t$,
\[{\cal Z}_t(\nu) = z_t(\nu;\beta_l,\mu_l)\,z_t(-\nu;\beta_r,\mu_r)\]
for some single function $z_t$, which is defined purely in terms of the chiral algebra $V$,
\[ 
z_t(\nu;\beta,\mu) = \sigma_{\beta,\mu}\big[e^{-\frc{i}2 \nu N_+ + i\nu\int_0^{t}dx\, \frak{j}(x)}\, e^{\frc{i}2 \nu N_+}\big].
\]

Assuming for a short while (but this will be proved in a few lines) that the following large-time limit exists, we define the charge transfer large deviation function $F$ by
\[
	F(\nu;\beta_{l,r},\mu_{l,r}):=\lim_{t\to\infty} \frc1t \log {\cal Z}_t(\nu)
\]
Due to the factorization of ${\cal Z}_t(u)$ we have:
\[
	F(\nu;\beta_{l,r},\mu_{l,r}) = f(\nu;\beta_l,\mu_l)+ f(-\nu;\beta_r,\mu_r),
\]
with $f(\nu;\beta,\mu):=\lim_{t\to\infty}\frac{1}{t} \log z_t(\nu;\beta,\mu)$. By scaling invariance, $f(\nu;\beta,\mu) = \beta^{-1}\, h(\nu;\beta\mu)$. We now claim 

\begin{propo} 
(1) The derivative of the chiral large deviation function $f$ is related to the current one-point function in the Gibbs state but at a shifted complex chemical potential. Explicitly
\beqa
	-i\frc{\p}{\p \nu} f(\nu;\beta,\mu) = \sigma_{\beta,\mu+i\nu\beta^{-1}}\big[ \frak{j}(0) \big], \label{dlambda}
\eeqa
(2) As a consequence, $f(\nu;\beta,\mu)=\pi[ (\beta\mu+i\nu)^2-(\beta\mu)^2]/(2\beta)$, for unitary theory.\\
(3) The charge transfer large deviation function is
\beq \label{F-u1}
F(\nu;\beta_{l,r},\mu_{l,r}) = \pi\frc{(\beta_l\mu_l+i\nu)^2 - (\beta_l\mu_l)^2}{2\beta_l}+ \pi\frc{(\beta_r\mu_r-i\nu)^2- (\beta_r\mu_r)^2}{2\beta_r},
\eeq
and the statistics of charge transfer is gaussian with mean $\pi{(\mu_l-\mu_r)}$ and covariance $\pi{(\beta^{-1}_l+\beta^{-1}_r)}$.
\end{propo}

The proof of this proposition is based on similar arguments as that for energy transfer plus an input from conformal field theory techniques which allows us to evaluate the current one-point function.

\medskip

{\bf Proof.}\\
First we take the derivative of $\log z_t(\nu;\beta,\mu)$ w.r.t. $i\nu$. This has two effects: it inserts $\int_0^tdx\, \frak{j}(x)$ inside the expectation value and it divides this expectation value by $z_t(\nu;\beta,\mu)$. By translation invariance, we may replace the large $t$ limit of the integral by the insertion of $t\, \frak{j}(t/2)$. The last factor $t$ combines with the factor $1/t$ in the definition of $f$, hence
\[ -i\partial_\nu f(\nu;\beta,\mu)=\lim_{t\to\infty} 
\frac{\sigma_{\beta,\mu}\big[e^{-\frc{i}2 \nu N_+ + i\nu\int_0^{t}dx\, \frak{j}(x)}\,\frak{j}(t/2)\, e^{+\frc{i}2 \nu N_+}\big]}{
\sigma_{\beta,\mu}\big[e^{-\frc{i}2 \nu N_+ + i\nu\int_0^{t}dx\, \frak{j}(x)}\, e^{+\frc{i}2 \nu N_+}\big]}.\]
Before taking the large time limit we manipulate this expression: using translation invariance we centred the inserted current at $0$ instead at $t/2$, and we then convert the integral $\int_{-t/2}^{t/2}dx\,\frak{j}(x)$ into $N_+-\big(\int_{t/2}^\infty+\int_{-\infty}^{-t/2}\big)dy\, \frak{j}(y)$. Recalling the definition of the Gibbs state $\sigma_{\beta,\mu}$, this yields to
\[ -i\partial_\nu f(\nu;\beta,\mu)=\lim_{t\to\infty} 
\frac{\sigma_{\beta,\mu+i\nu\beta^{-1}}\big[e^{-\frc{i}2 \nu N_+}\,e^{+\frc{i}2 \nu N_+ - i\nu\big(\int_{t/2}^\infty+\int_{-\infty}^{-t/2}\big)dy\, \frak{j}(y)}\,\frak{j}(0)\big]}{\sigma_{\beta,\mu+i\nu\beta^{-1}}\big[e^{-\frc{i}2 \nu N_+}\,e^{+\frc{i}2 \nu N_+ - i\nu\big(\int_{t/2}^\infty+\int_{-\infty}^{-t/2}\big)dy\, \frak{j}(y)}\big]}.\]
Now, as for energy transfer, the perturbative expansion  of this ratio in a power series of $\nu$ involves connected correlation functions of products of the current operator $\frak{j}(0)$ times series of multiple ordered integrals of $\frak{j}(y)$. Since the currents $\frak{j}(y)$ are located at distance at least $t/2$ from the origin, the clustering property of connected correlation functions ensures that we can safely take the large $t$ limit. All contributions go to zero except the zero order in the perturbative expansion, and this amounts to take directly the large $t$ limit in the above equation. Thus
\[ -i\partial_\nu f(\nu;\beta,\mu)=\sigma_{\beta,\mu+i\nu\beta^{-1}}\big[\frak{j}(0)\big].\]
Actually,  the proof can be slightly simplified using the fact that $N_+$ commutes with the current operator $\frak{j}(y)$ in the pure $u(1)$  case that we are presently dealing with, but we need to keep this more general argument for next section.

We proved point (1) of the proposition. To prove point (2) we need to compute $\sigma_{\beta,\mu}\big[\frak{j}(0)\big]$ and then shifts $\beta\mu$ into $\beta\mu+i\nu$. This is done using conformal field theory techniques and modular transformation. Let $\chi[\tau,z]$ be affine $u(1)$ character: $\chi[\tau,z] = \Tr\big( e^{2\pi i \tau(L_0-c/24)} e^{4\pi i z j_0}\big)$ with normalization $\frak{j}(z) \frak{j}(0)\sim 1/(z^2)$. Since according to Section \ref{Sect:Gibbs}, $H_+= \frac{2\pi}{R} (L_0-c/24)$ at finite size, we need to set $\tau = i\beta/R$ and $z = -i\beta \mu/2$, so that
\[ 
\sigma_{\beta,\mu}\big[\frak{j}(0)\big]= 
\lim_{R\to\infty} \frac{1}{\beta R} \frac{\partial}{\partial\mu} \log \chi[i\beta/R,-i\beta \mu/2], 
\]
by translation invariance. To evaluate the large size limit we implement a modular transformation. Modular property of characters of affine algebras are well known. They form representations of the modular group, and we may write 
\[
	\chi[\tau,z]= e^{-i2\pi{z^2}/{\tau}}\, \sum_p S_p\, \chi_p[-1/\tau,-z/\tau].
\]
where the sum is over a set of affine $u(1)$ modules, including the identity module $V$, and $S_p$ are coefficient of the so-called modular $S$-matrix \cite{KacWakimoto}. For unitary theory, the leading contribution for $R$ large comes from the identity module in the sum. This gives
\[ \sigma_{\beta,\mu}\big[\frak{j}(0)\big] = \pi\,\mu.\]
Integrating $-i\partial_\nu f= \sigma_{\beta,\mu+i\nu\beta^{-1}}\big[\frak{j}(0)\big]$ yields the formula for $f$.

\section{Energy and charge flow statistics and its large deviation function}\label{sectboth}

We now look at the large deviation function combining energy and charge transfers associated to the two {\it commuting} observables $E :=\frac{1}{2}(H_o^r-H_o^l)$ and $Q:=\frac{1}{2}(N_o^r-N_o^l)$. As above we have to compute the generating function
${\cal Z}_t(\lambda,\nu) := \omega_{\rm stat}\big[ e^{i\nu Q(t)}\, e^{i\lambda E(t)}\, e^{-i\nu Q}\, e^{-i\lambda E} \big]$,
and the large deviation function
\[ F(\lambda,\nu;\beta_{l,r},\mu_{l,r}):=\lim_{t\to\infty} \frac{1}{t} \log {\cal Z}_t(\lambda,\nu).\]
As before, ${\cal Z}_t$ factorizes in left/right expectations so that the large deviation function reads
\beq \label{FullEQ}
F(\lambda,\nu;\beta_{l,r},\mu_{l,r})= f(\lambda,\nu;\beta_l,\mu_l)+  f(-\lambda,-\nu;\beta_r,\mu_r).
\eeq
The function $ f(\lambda,\nu;\beta,\mu)$ is the chiral large deviation function which is computed using data from the chiral algebra $V$ only:
\[ f(\lambda,\nu;\beta,\mu)=\lim_{t\to\infty} \frac{1}{t}\log\,
\sigma_{\beta,\mu}\big[{\cal O}^t_q\,e^{\frc{i}2 \nu N_+} \cdot {\cal O}^t_e\,e^{\frc{i}2 \lambda H_+}\big] \]
with $\sigma_{\beta,\mu}$ the Gibbs state over $V$ at temperature $1/\beta$ and chemical potential $\mu$, and 
\[ {\cal O}^t_e=e^{-\frc{i}2 \lambda H_+ + i\lambda \int_0^tdx\,\frak{h}(x)},\quad
{\cal O}^t_q=e^{-\frc{i}2 \nu N_+ + i\nu \int_0^tdx\,\frak{j}(x)} .\]

Let us now compute $f$. We follow the same strategy as in previous section, and we shall be more brief.
As before, we take derivatives w.r.t.~$\lambda$ and $\nu$. Using translation invariance to replace the integrals of the energy or charge densities by the insertion of the corresponding operator at the middle point $t/2$, we get
\begin{eqnarray*}
-i\partial_\lambda  f(\lambda,\nu;\beta,\mu)=\lim_{t\to\infty} 
\frac{\sigma_{\beta,\mu}\big[ {\cal O}^t_q\, e^{\frc{i}2 \nu N_+} \cdot {\cal O}^t_e\ \frak{h}(t/2)\, e^{\frc{i}2 \lambda H_+}\big]}{
\sigma_{\beta,\mu}\big[{\cal O}^t_q\, e^{\frc{i}2 \nu N_+} \cdot {\cal O}^t_e\, e^{\frc{i}2 \lambda H_+} \big]},\\
-i\partial_\nu  f(\lambda,\nu;\beta,\mu)=\lim_{t\to\infty} 
\frac{ \sigma_{\beta,\mu}\big[ {\cal O}^t_q\ \frak{j}(t/2)\, e^{\frc{i}2 \nu N_+} \cdot {\cal O}^t_e\, e^{\frc{i}2 \lambda H_+} \big]}{
\sigma_{\beta,\mu}\big[{\cal O}^t_q\, e^{\frc{i}2 \nu N_+} \cdot {\cal O}^t_e\, e^{+\frc{i}2 \lambda H_+} \big]}.
\end{eqnarray*}
This ratio of expectations only involve connected correlations functions. So, by the usual argument this garanties that we can take the naive large time limit which amounts to replace ${\cal O}^t_e$ by $e^{\frac{i}{2} \lambda H_+}$ and ${\cal O}^t_q$ by $e^{\frac{i}{2} \nu N_+}$. As a consequence, we get

\begin{propo} (1) We have:
\begin{eqnarray}\label{dfEQa}
-i\partial_\lambda  f(\lambda,\nu;\beta,\mu)&=&\sigma_{\beta-i\lambda,\frac{\beta\mu+i\nu}{\beta-i\lambda}}\big[\frak{h}(0)\big],\\
-i\partial_\nu  f(\lambda,\nu;\beta,\mu)&=& \sigma_{\beta-i\lambda,\frac{\beta\mu+i\nu}{\beta-i\lambda}}\big[ \frak{j}(0) \big].
\label{dfEQb}
\end{eqnarray}
(2) For unitary theory, we have
\begin{eqnarray} \label{fEQ}
 f(\lambda,\nu;\beta,\mu)= \frac{c\pi}{12}\big(\frac{1}{\beta-i\lambda}-\frac{1}{\beta}\big) 
 + \frac{\pi(\beta\mu+i\nu)^2}{2(\beta-i\lambda)} - \frac{\pi \beta\mu^2}{2}.
 \end{eqnarray}
 The large deviation function for energy and charge transfers is given by:
  $F(\lambda,\nu;\beta_{l,r},\mu_{l,r})= f(\lambda,\nu;\beta_l,\mu_l)+  f(-\lambda,-\nu;\beta_r,\mu_r)$.
\end{propo}

{\bf Proof.}\\
We already argued for point (1). To get point (2) we have to evaluate both one point functions (\ref{dfEQa},\ref{dfEQb}). This is done as previously, using modular property of characters and $\partial_\beta\log\chi=-R\,\sigma_{\beta,\mu}\big[ \frak{h}(0)-\mu \frak{j}(0)\big]$ and $\partial_\mu\log\chi=\beta R\, \sigma_{\beta,\mu}\big[ \frak{j}(0) \big]$. For unitary theory, we get:
\begin{eqnarray} \label{1ptbetamu}
\sigma_{\beta,\mu}\big[ \frak{j}(0)\big]= \pi{\mu} ,\quad
\sigma_{\beta,\mu}\big[ \frak{h}(0) \big]=\frac{c\pi}{12\beta^2} + \frac{\pi\mu^2}{2}.
\end{eqnarray}
Notice the extra term in $\sigma_{\beta,\mu}\big[\frak{h}(0) \big]$ for non zero chemical potential $\mu$. This terms ensures that the two differential equations (\ref{dfEQa},\ref{dfEQb}) are compatible. Integrating them with boundary condition $f(0,0;\beta,\mu)=0$ yields formula (\ref{fEQ}).

Formula (\ref{fEQ}) indicates that charge and energy flows are not statistically independent, even at large time, and this is a direct consequence of eq.(\ref{1ptbetamu}) for the one-point chiral energy density at nonzero chemical potential. As a consequence, at identical temperatures but different chemical potentials, the mean energy current does not vanish 
$\langle J_E \rangle_{\Delta\beta=0,\Delta\mu\not=0} = {\pi}(\mu_l^2-\mu_r^2)$.

\section{Conclusion and discussion}\label{sectdisc}

We have constructed non-equilibrium steady states in conformal field theories and derived the associated  large deviation functions. It is worth noticing the universal character of these states. As a consequence the large deviation functions for energy and charge transfers are very universal, depending only on universal constant ($\hbar,\ k_B$) and on the temperatures $T_{l,r}$ and chemical potentials $\mu_{l,r}$. Although we used a conformal field theory approach, we provided arguments in favour of the universality of our construction in the low energy regime. Remark also that we would have got identical steady states if we would have prepared the two infinite left/right subsystems at different temperatures and connected them with some finite-length subsystem initially in another distribution. It should also be noticed that naive Fourier's law for the energy current as a function of the temperature gradient is broken in our model \footnote{In low dimension one naturally expects a breakdown of the Fourier's law with energy current scaling as $L^{-\alpha}$ for some exponent $\alpha$ if the two reservoirs are at a distance $L$ apart.}. Indeed, in our model, the temperature profile, say specified by the local energy density, is flat, although the energy current is non zero (this may be related to the zooming prescription that we alluded to in Section \ref{sectresults}). As we shall explain \cite{BDnofutur}, the key equations (\ref{dfEQa},\ref{dfEQb}) relating derivatives of the large deviation functions to the linear response theory but at shifted temperatures and chemical potentials fit into consequences of PT-symmetry, as does the reflection-less Lesovik-Levitov formula for charge transfer in mesoscopic free electronic systems. Generalisations of our construction to include non-topological defects, or massive integrable perturbations, or to higher dimensions should provide valuable information on non-equilibrium quantum physics.
\vskip 0.5 truecm

{\bf Acknowledgements}: We thank J. Cardy and M. J. Bhaseen for interesting discussions. This work was in part supported by ANR contract ANR-2010-BLANC-0414.

\appendix

\section{Two-time energy measurement in the steady state}\label{appu}

Here we briefly present the derivation of the large deviation function according to the measurement scheme described by eqs.(\ref{ZtZtu},\ref{Zt}). According to (\ref{Zt}), take
\beq\label{aor}
	\Or' =  e^{-i\lt(\frc\lambda2-u\rt) E} e^{i\lambda (E+\Delta E(t))}e^{-i\lt(\frc\lambda2+u\rt)E}.
\eeq
We follow the lines of reasoning of Section \ref{propoFCS}. Defining $\Delta E(t,a)$ as in (\ref{DeltaEa}) and using the BCH formula, we may represent the regularization of $\Or'$ by
\beq\label{aOrexact}
	\Or'\mapsto \Or_{a}(\lambda,u):=
	U_{u-\lambda/2}^E\lt(\Pexp \int_0^\lambda dz \lt(
	i\,U_z^E\lt(\Delta E(t,a)\rt)\rt)\rt)
\eeq
with $U_t^E$ defined in (\ref{UtE}). In order to describe the two-time measurement, recall that we need to integrate over the variable $u$, see (\ref{ZtZtu}). In the infinite-length limit, the energy spectrum is continuous, hence the integration range is $(-\tau,\tau)$ with $\tau\to\infty$. The precise proportionality factor in (\ref{ZtZtu}) is not important, except for the fact that it becomes proportional to $1/\tau$ as $\tau\to\infty$. Hence, our starting expressions are
\beq \label{ainit}
	\t F(\lambda) = \lim_{t\to\infty} t^{-1}
	\log \lt(\lim_{\tau\to\infty}
	\tau^{-1} \int_{-\tau}^\tau du\,
	\omega_{\rm stat}\big[\Or_{\ep,a}(\lambda,u) \big]\rt)
\eeq
The main point is to notice that the integrand inside the $u$ integral in the first equation of (\ref{init}) does not vanish as $|u|\to\infty$, but rather converges to a constant. Hence, the $\tau\to\infty$ limit is given by this constant. In order to evaluate this constant, notice that
\beqa
	\lefteqn{U_{u-\lambda/2+z}^E(\Delta E(t,a))} \label{largeu} \\
	&=& \lt\{\ba{ll}\displaystyle
	\int_a^t dx\,\lt(h_-\lt(x-\frc u2+\frc\lambda4-\frc z2\rt) - h_-\lt(x+\frc u2-\frc \lambda4+\frc z2\rt)\rt)
	& \mbox{$u$ large and positive}
	\z
	\int_a^t dx\,\lt(h_+\lt(-x+\frc u2-\frc \lambda4+\frc z2\rt) - h_+\lt(-x-\frc u2+\frc \lambda4-\frc z2\rt)\rt)	
	& \mbox{$u$ large and negative.}
	\ea\rt. \no
\eeqa
We then use
\[
	\Or_{a}(\lambda,u):=
	\lt(\Pexp \int_0^\lambda dz\lt(
	i\,U_{u-\lambda/2+z}^E
	\lt(\Delta E(t,a)\rt)\rt)\rt)
\]
and for large $|u|$, we observe that both cases $u>0$ and $u<0$ in (\ref{largeu}) involve two terms largely separated, so that we can make use of the factorization property (\ref{factori}) of $\omega_{\rm stat} = \sigma_{\beta_l,\beta_r}$. Using translation invariance the $u$-dependence of the result disappears. We can also use translation invariance to take away the explicit $\lambda$-dependence in the arguments of $h_\pm$, and then we may use the change of variable $x\mapsto t+a-x$ followed by a translation by $\pm(t+a)$ in order to change the sign of $x$. This allows to bring each factor to a form involving the path-ordered exponential in (\ref{fbl}). We then find
\beq
	\lim_{\tau\to\infty}
	\tau^{-1} \int_{-\tau}^\tau du\,
	\omega_{\rm stat}\big[\Or_{\ep,a}(\lambda,u) \big]
	\sim e^{t(f(\lambda;\beta_r)+f(-\lambda;\beta_r))} + e^{t(f(\lambda;\beta_l)+f(-\lambda;\beta_l))}\quad
	\mbox{as}\quad t\to\infty.
\eeq
This shows that the difference between $\t F(\lambda)$ and $F(\lambda)$ is nonzero. This indicates that the two-time measurement directly in the steady state, in CFT, does not correspond to the same physics as the two-time measurement where the first measurement is made at the connection point.


\begin{thebibliography}{}

\bibitem{AB} W. H. Aschbacher and  J.-M. Barbaroux, ``Out of equilibrium correlations in the XY chain'', Lett. Math. Phys. 77 (2006), 11-20.

\bibitem{AschPill} W.H. Aschbacher and C.-A. Pillet, ``Non-Equilibrium Steady States of the XY Chain", J. Stat. Phys. 112 (2003) 1153.

\bibitem{BD11} D. Bernard, B. Doyon, ``Energy flow in non-equilibrium conformal field theory.", J. Phys. A 45, 362001 (2012).

\bibitem{BD-RLM} D. Bernard, B. Doyon, ``Full counting statistics in the resonant-level model.", J. Math. Phys. 53 (2012)122302.


\bibitem{BDnofutur} D. Bernard, B. Doyon, ``PT-symmetry and symmetry classes of quantum full counting statistics", in preparation.

\bibitem{BLR} F. Bonetto, J.L. Lebowitz, L. Rey-Bellet, ``Fourier's Law: a Challenge for Theorists'', in Mathematical Physics 2000, Imp. Coll. Press, London, 2000, pp. 128-150; arXiv:math-ph/0002052.

\bibitem{CalabCardy} P. Calabrese and J. Cardy, ``Time-dependence of correlation functions following a quantum quench", Phys. Rev. Lett. 96, 136801 (2006); "Quantum quenches in extended systems", J. Stat. Mech. (2007) P06008.

\bibitem{CalabEssler} P. Calabrese, F.H.L. Essler and M. Fagotti, ``Quantum Quench in the Transverse Field Ising Chain", Phys. Rev. Lett. 106, 227203 (2011).

\bibitem{Cardy_bdry} J. Cardy, ``Conformal Invariance and Surface Critical Behavior", Nucl. Phys. B240 [FS12], 514-532, 1984.

\bibitem{CardyStef} J. Cardy, ``The Ubiquitous 'c': from the Stefan-Boltzmann Law to Quantum Information'', J. Stat. Mech. 1010:P10004, 2010.

\bibitem{Cald-Legg} A.O. Caldeira and A.J. Leggett, ``Influence of dissipation on quantum tunneling in macroscopic systems", Phys. Rev. Lett., 46, (1981) 211.

\bibitem{Cast95} H. Castela, X. Zotos, P. Prelovsek, ``Integrability and Ideal Conductance at Finite Temperatures", Phys. Rev. Lett. {\bf 74} (1995) 972.

\bibitem{CohenGal} G. Gallavotti and E. Cohen, ``Dynamical ensembles in non-equilibrium statistical mechanics", Phys. Rev. Lett. 74 (1995) 2694.

\bibitem{Davies} E.B. Davies, ``Markovian master equations'', Commun. Math. Phys 39 (1974) 91-110.

\bibitem{DHouches} B. Doyon, ``Nonequilibrium density matrix for thermal transport in quantum field theory'',  arXiv:1212.1077.

\bibitem{Espo} M. Esposito, U. Harbola and S. Mukamel, ``Nonequilibrium fluctuations, fluctuation
theorems and counting statistics in quantum systems", Rev. Mod. Phys. 81 (2009) 16651702.

\bibitem{vertex-op} I. Frenkel, J. Lepowsky, A. Meurman, ``Vertex operator algebras and the monster", Academic Press 1989. 

\bibitem{Giama92} T. Giamarchi, A.J. Millis, ``Conductivity of a Luttinger liquid", Phys. Rev. {\bf B46} (1922) 9325.

\bibitem{GKLH} A. O. Gogolin, R. M. Konik, A. W. W. Ludwig, H. Saleur, ``Counting statistics for the Anderson impurity model: Bethe ansatz and Fermi liquid study'', Ann. Phys. (Leipzig) 16, 678 (2007).

\bibitem{GGM} D. B. Gutman, Yu. Gefen and A. D. Mirlin, ``Full counting statistics of Luttinger liquid conductor''m Phys. Rev. Lett. 105, 256802 (2010).

\bibitem{KacWakimoto} V.G. Kac, M. Wakimoto, ``Modular and conformal invariance constraints in representation theory of affine Lie algebras", Adv. Math. 70 (1988) 156.

\bibitem{Karevski} D. Karevski, V. Popkov, and G.M. Sch¬utz, ``Exact matrix product solution for the boundary-driven Lindblad XXZ-chain", arXiv:1211.7010.

\bibitem{Moore12}  C. Karrasch R. Ilan and J. E. Moore, ``Nonequilibrium thermal transport and its relation to linear response", arXiv:1211.2236.

\bibitem{xxx} I. Klich, ``Full counting statistics: an elementary derivation of Levitov's formula", arXiv:cond-mat/0209642.\\
K. Schonhamer, ``Full counting statistics for non-interacting fermions: exact results and the Levitov-Lesovik formula", Phys. Rev. B 75 (2007) 2053229.\\
J.E. Avron, S. Bachmann, G.M. Graf and I. Klich, ``Fredholm determinants and the statistics of charge transport", Commun. Math. Phys. 280 (2008) 807-829.

\bibitem{KS} A. Komnik and H. Saleur, ``Full counting statistics of chiral Luttinger liquids with impurities", Phys. Rev. Lett. 96, 216406 (2006).

\bibitem{InfinitReserv} J. Lebowitz and H. Spohn, ``Irreversible themodynamics for quantum systems weakly coupled to thermal reservoirs", Adv. Chem. Phys. 39 (1978) 109-142.

\bibitem{LLformula} L.S. Levitov and G.B. Lesovik, ``Charge distribution in quantum shot noise", JETP Lett. 58 (1993) 230-235; "Quantum measurement in electric circuit", arXiv:cond-mat/9401004.

\bibitem{Lindblad} G. Lindblad, ``On the generators of quantum dynamical semigroups", Commun. Math. Phys. 48 (1976) 119-130.

\bibitem{vertex-op-cyl} G. Mason and M. P. Tuite, ``Vertex Operators and Modular Forms'', in: A Window into Zeta and Modular Physics, ed Kirsten, K. and Williams, F., MSRI Publications 57 (2010), 183--278, C.U.P.

\bibitem{Mintchev} M. Mintchev and P. Sorba, ``Luttinger Liquid in Non-equilibrium Steady State", J. Phys. A: Math. Theor. 46 (2013) 095006.

\bibitem{JBZ} V.B. Petkova, J.-B. Zuber, ``Generalised twisted partition functions", Phys. Lett. B 504 (2001) 533.

\bibitem{Pros11} T. Prosen, ``Open XXZ Spin Chain: Nonequilibrium Steady State and a Strict Bound on Ballistic Transport", Phys. Rev. Lett. {\bf 106} (2011) 217206.

\bibitem{Prosen} T. Prosen, ``Exact Nonequilibrium Steady State of a Strongly Driven Open XXZ Chain", Phys. Rev. Lett. 107 (2011) 137201.

\bibitem{Ruelle} D. Ruelle, ``Natural non-equilibrium states in quantum statistical mechanics", J. Stat. Phys. 98 (2000) 57.

\bibitem{OscilChain} K. Saito and A. Dhar, ``Fluctuation Theorem in Quantum Heat Conduction", Phys. Rev. Lett. 99 (2007) 180601.

\bibitem{Aff05} J. Sirker, R.G. Pereira, I. Affleck, ``Diffusion and Ballistic Transport in One Dimensional Quantum Systems", Phys. Rev. Lett. {\bf 103} (2009) 216602.

\bibitem{SpohnLebo} H. Spohn and J.L. Lebowitz, ``Stationary Non-Equilibrium States of Infinite Harmonic Systems", Commun. Math. Phys., 54 (97), 1977.


\bibitem{extensions} In preparation by various collaborations including Y. Chen, D. Bernard, B. Doyon, M. Hoogeveen, A. De Luca and J. Viti.





\end{thebibliography}
\end{document}